\documentclass[a4paper,fleqn]{cas-sc}
\usepackage{amsfonts}
\usepackage{array} 

\usepackage{amsmath}
\usepackage{amssymb}
\usepackage{graphicx}
\usepackage[square,numbers,sort&compress]{natbib}
\usepackage{tabularx}
\usepackage{latexsym}
\usepackage[misc]{ifsym}
\RequirePackage{fix-cm}
\usepackage{subcaption}
\usepackage{longtable}
\usepackage{caption}
\usepackage[utf8]{inputenc}
\usepackage{textcomp}
\usepackage{colortbl}
\usepackage{xcolor}
\definecolor{lightblue}{rgb}{0.8, 0.9, 1.0}

\def\tsc#1{\csdef{#1}{\textsc{\lowercase{#1}}\xspace}}
\tsc{WGM}
\tsc{QE}
\tsc{EP}
\tsc{PMS}
\tsc{BEC}
\tsc{DE}

\ExplSyntaxOn 
\cs_gset:Npn \__first_footerline: 
{ \group_begin: \small \sffamily \__short_authors: \group_end: } \ExplSyntaxOff

\usepackage{longtable}
\begin{document}
	\let\WriteBookmarks\relax
	\title[mode = title]{Toward Realistic Cinema: The State of the Art in Mechatronics for Modern Animatronics} 
	\shortauthors{Riham Hilal, et.al.}
	\shorttitle{Mechatronics \& Animatronics}
	\author[1,3]{Riham M. Hilal}
	\ead{riham.hilal@ejust.edu.eg, riham.hilal@alexu.edu.eg} %
	\address[1]{Art \& Design Programs, Egypt-Japan University of Science and Technology E-JUST,\\
		New Borg El-Arab City, Alexandria 21934, Egypt.\\}
	
\author[2]{Haitham El-Hussieny}
\ead{haitham.elhussieny@ejust.edu.eg} %
	\author[2]{Ayman A. Nada}
	\ead{ayman.nada@ejust.edu.eg} %
	\address[2]{Mechatronics and Robotics Engineering, School of Innovative Design Engineering,\\
		Egypt-Japan University of Science and Technology E-JUST,\\
		New Borg El-Arab City, Alexandria 21934, Egypt.\\}
	\address[3]{Faculty of Fine Arts, Alexandria University, Alexandria 21526, Egypt.\\}

	\begin{abstract}
The pursuit of realism in cinema has driven significant advancements in animatronics, where the integration of mechatronics, a multidisciplinary field that combines mechanical engineering, electronics, and computer science, plays a pivotal role in enhancing the functionality and realism of animatronics. This interdisciplinary approach facilitates smoother characters movements and enhances the sophistication of behaviors in animatronic creatures, thereby increasing their realism. This article examines the most recent developments in mechatronic technology and their significant impact on the art and engineering of animatronics in the filmmaking. It explores the sophisticated integration of system components and analyzes how these enhancements foster complexity and integration, crucial for achieving unprecedented levels of realism in modern cinema. Further, the article delves into in-depth case studies of well-known movie characters, demonstrating the practical applicability of these state-of-the-art mechatronic solutions in creating compelling, lifelike cinematic experiences. This paper aims to bridge the gap between the technical aspects of mechatronics and the creative demands of the film industry, ultimately contributing to the ongoing evolution of cinematic realism.
	\end{abstract}
	

	\maketitle
\section{Introduction}
Prior to the widespread adoption of mechatronic technology, the film industry faced numerous mechanical constraints that significantly affected production efficiency and creative possibilities. The integration of mechatronics—an interdisciplinary field combining mechanical engineering, electronics, computer science, and control engineering—has transformed these challenges into opportunities for innovation. Historically, the film industry relied on traditional mechanical systems for camera operation, lighting, and set design. These systems often posed limitations such as rigidity and lack of flexibility \cite{ahern2018cinema}, labor-intensive processes \cite{alabi2019challenges}, precision limitations \cite{ahern2018cinema} and the lack of integration between different mechanical systems made it difficult to synchronize various elements of production, such as camera movements with lighting changes or special effects.

 These challenges restricted both the art framework for creativity and the technical capabilities of cinema production, as well as limited the feasible speeds of old film copying methods. These earlier mechanisms, which mainly relied on mechanical gears and sprockets, frequently experienced problems that weakened the quality of the final film production \cite{Stoker2000}. Nowadays, mechatronics plays a crucial role in scenes that combine practical effects with computer-generated imagery (CGI). Animatronics provide a physical presence that can be enhanced with CGI, giving film-makers more flexibility to achieve visual effects that were previously impossible \cite{Stoker2000}. For instance, the subtle movements of facial features or the fluid motion of limbs are much more refined with mechatronic technology. Traditional animatronics relied heavily on simple mechanics like pneumatics and hydraulics, which limited their range of motion. Mechatronics allows for the incorporation of advanced robotics into creature design, which can simulate muscle movement and facial expressions more naturally and with greater control. With advancements in mechatronics, designers can program creatures to perform complex sequences of actions that can be adjusted and repeated with high precision. This programmability also enhances the safety and repeatability of complex scenes, reducing the need for retakes and decreasing production costs.
 
 Mechatronic creatures can interact more effectively with human actors and environments due to sensors and real-time processing capabilities. These features allow the creatures to respond to external stimuli, adding an unscripted element to interactions that can enhance the believability of scenes. Overall, mechatronics has enabled film-makers to push the boundaries of what is visually possible, creating more engaging and believable non-human characters that captivate audiences. In general, scholarly publications about animatronics in motion pictures enhance our comprehension of the convergence of technology, art, and narrative in the motion picture business.
 
 This article explores novel ways to use animatronics to animate movie characters through innovative processes. It examines the most recent developments in mechatronic technology and their significant impact on the art and engineering of animatronics in film-making. The paper aims to explore the sophisticated integration of system components and analyze how these enhancements foster complexity and integration, crucial for achieving unprecedented levels of realism in modern cinema. Additionally, the article delves into in-depth case studies of well-known movie characters, demonstrating the practical applicability of these state-of-the-art mechatronic solutions in creating compelling, lifelike cinematic experiences. 
 
 This paper is organized as follows: Section 2 provides the theoretical and technical aspects of mechatronics with practical applications in the film industry, offering a comprehensive overview of how these systems are integral to the evolution of cinematic technology and artistry. Section 3 showcases notable examples of animatronics in films, demonstrating the diverse applications of this technology across different models such as humans, sharks, bees, bats, monsters, and vampires. Section 4 reviews the latest technological advancements in mechatronics relevant to film-making and analyzes how these advancements have facilitated smoother and more sophisticated movements in animatronic creatures. Section 5 explores the significant impact of mechatronics on the design and realism of cinematic creatures, highlighting the technological advancements that have enabled more dynamic and interactive performances. Section 6 discusses the future of creature animation, speculating on future trends and potential innovations in mechatronics that could further transform film-making. Section 7 provides in-depth case studies of notable movie characters enhanced through mechatronics, demonstrating the practical application and specific mechatronic solutions used to achieve lifelike effects. Finally, Section 8 summarizes the key points discussed, reflecting on the transformational role of mechatronics in animatronics and its broader implications for the film industry.
	
	\begin{figure}[tb]
	\noindent 
	\begin{minipage}[tb]{0.5\textwidth}
		\begin{center}
			\includegraphics[width=0.80\textwidth]
			{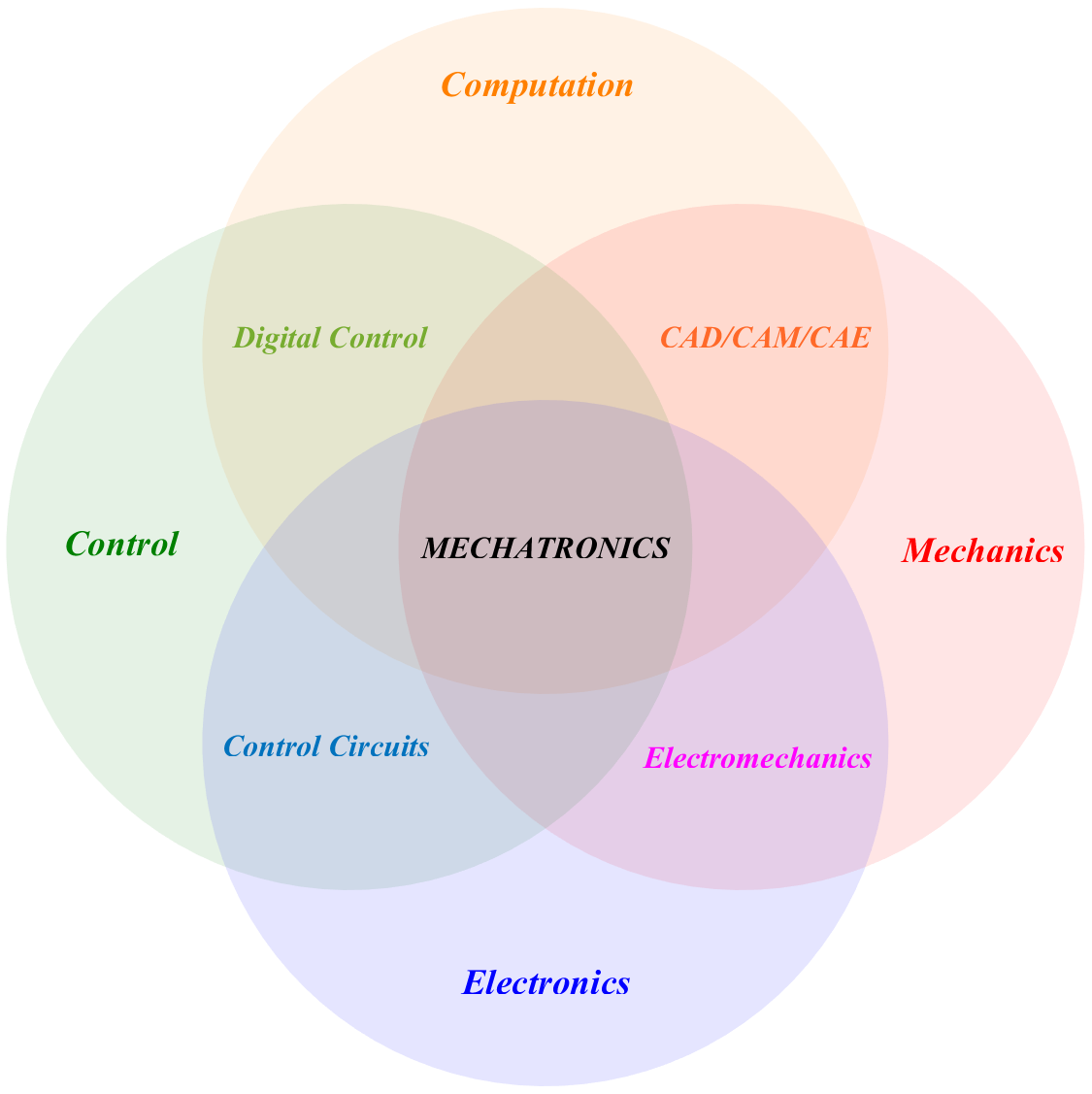}
			\caption{Integrative Framework of Mechatronics}
			\label{Fig: Mechatronics1}
		\end{center}
	\end{minipage}\noindent 
	\begin{minipage}[tb]{0.50\textwidth}
		\begin{center}
			\includegraphics[width=0.95\textwidth]
			{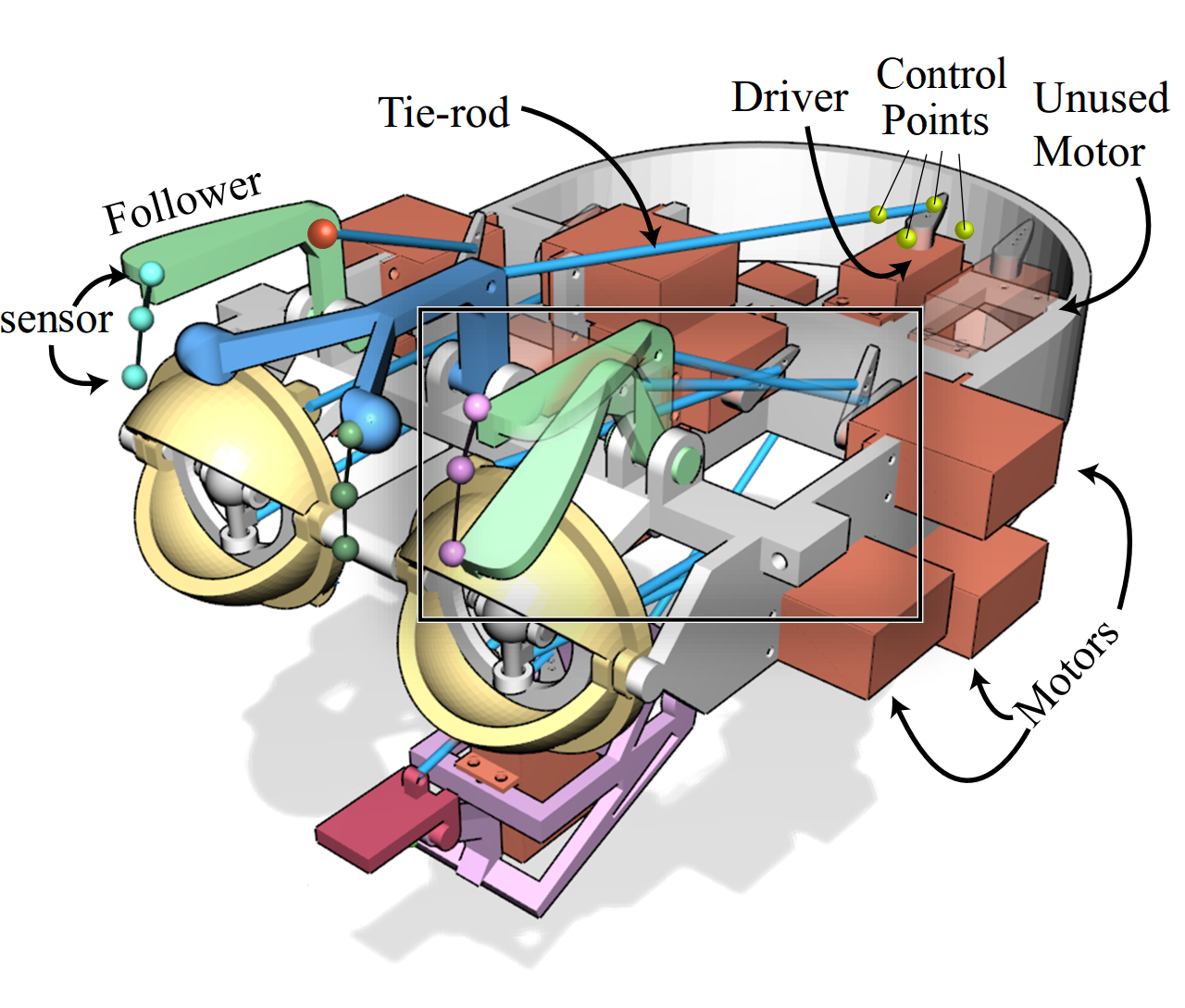}
			\caption{Animatronic Face Mechanism, \citet{Huber2021}}
			\label{Fig: Mechatronics2}
		\end{center}
	\end{minipage}\newline
\end{figure}
\section{The Essence of Mechatronics in Cinematic Design}
This section provides the theoretical and technical aspects of mechatronics with practical applications in the film industry, providing a comprehensive overview of how these systems are integral to the evolution of cinematic technology and artistry. As seen in Fig.(\ref{Fig: Mechatronics1}), Mechatronics is an interdisciplinary field that plays a major role in modern cinema, namely in the creation of advanced animatronic characters and machines. This study examines the impact of computational technologies, specifically CAD/CAM/CAE (Computer-Aided Design, Computer-Aided Manufacturing, and Computer-Aided Engineering), on the advancement of animatronics. It also investigates the utilization of control systems and digital controls to achieve exact manipulation for the purpose of creating desired cinematic effects. The subject encompasses the technical aspects of physical design and material science that guarantee lifelike movement and structural stability, as well as the importance of electronics and electromechanics in incorporating sensors for motion and responsiveness. This section delves into the design of control circuits that combine electronics and control systems. It also explores how the merging of these fields in mechatronics results in creative cinematic solutions. The section provides examples of specific advancements in animatronics that demonstrate this integration.

In this context, Fig.(\ref{Fig: Mechatronics2}) depicts a clear mechatronic system, specifically designed for animatronics, integrating mechanical components, electronics, and control systems. This mechatronic system combines electronics (motors and maybe sensors as suggested by control arrows), mechanical components (tie-rods, followers, drivers), and control systems (control points, unused motors suggesting modular architecture). It designed the system for synchronized tasks in animatronics, prioritising accuracy, dependability, and consistent movements. Every component is purposefully placed and designed to improve system functioning and performance, demonstrating mechatronics' interdisciplinary nature. This field develops technologically advanced and flexible solutions using mechanical, electronic, and systems control concepts.

\begin{figure}[tb]
	\noindent 
	\begin{minipage}[tb]{0.28\textwidth}
		\begin{center}
			\includegraphics[width=0.90\textwidth]
			{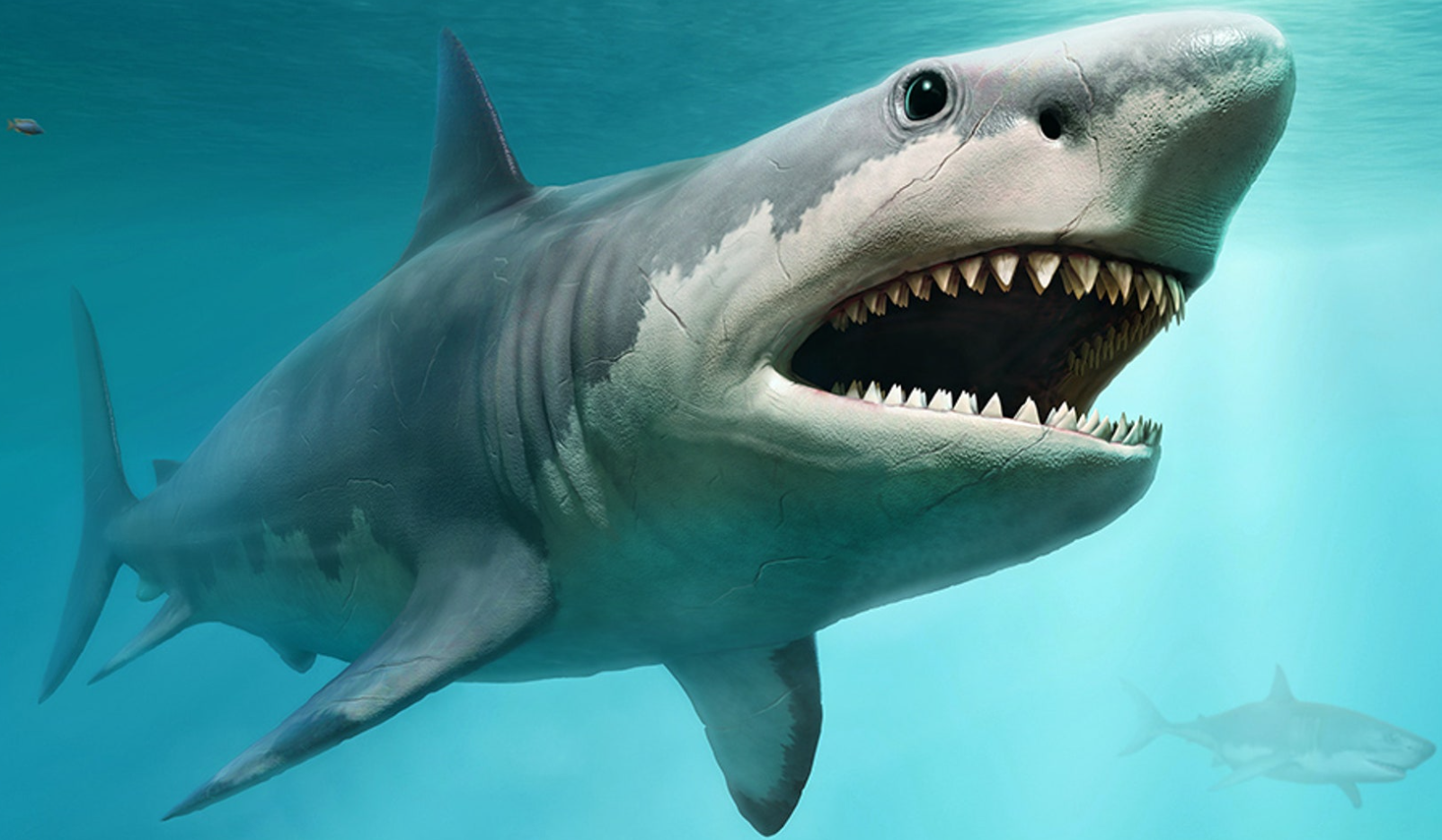}
			\caption{Mackerel Shark}
			\label{Fig: Shark1}
		\end{center}
	\end{minipage}\noindent 
	\begin{minipage}[tb]{0.33\textwidth}
		\begin{center}
			\includegraphics[width=0.92\textwidth]
			{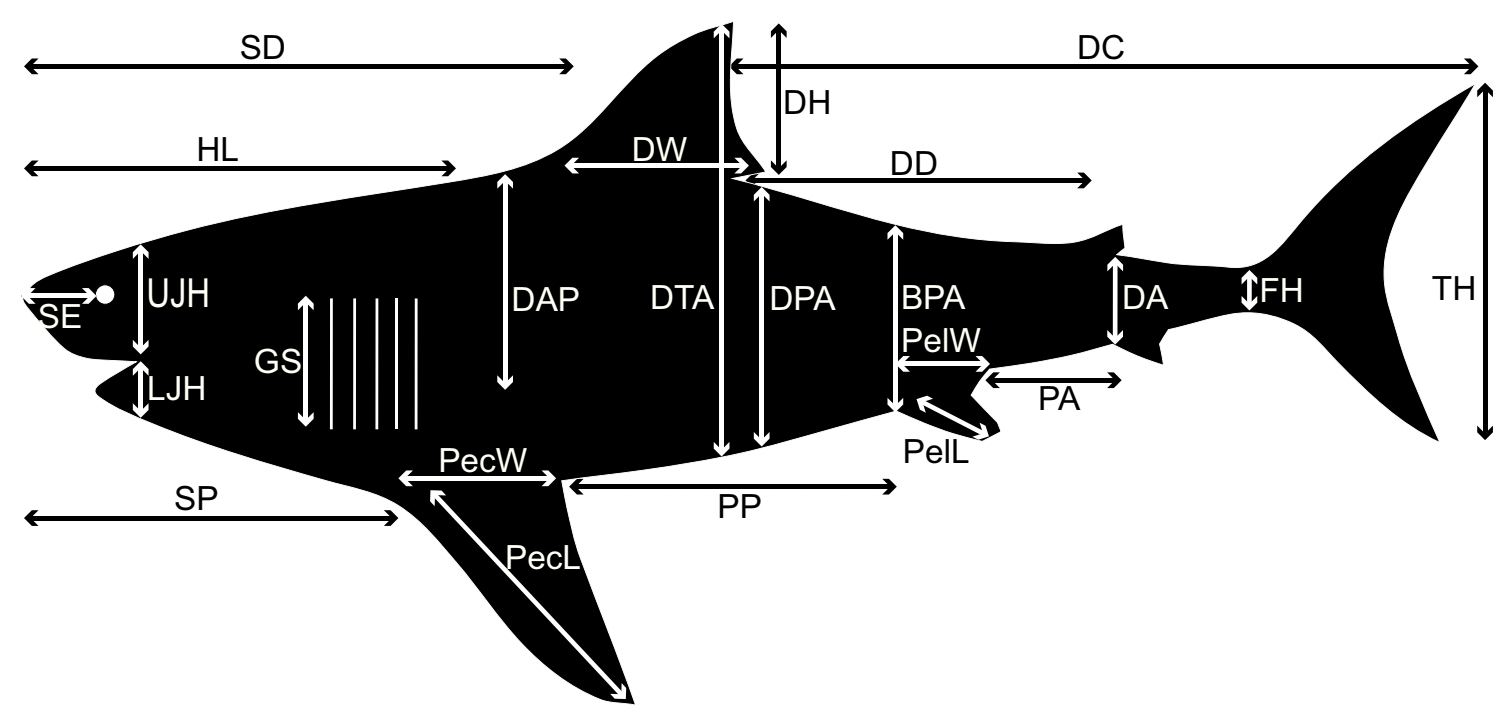}
			\caption{Geometric Dimensions}
			\label{Fig: Shark2}
		\end{center}
	\end{minipage}
	\begin{minipage}[tb]{0.35\textwidth}
		\begin{center}
			\includegraphics[width=0.99\textwidth]
			{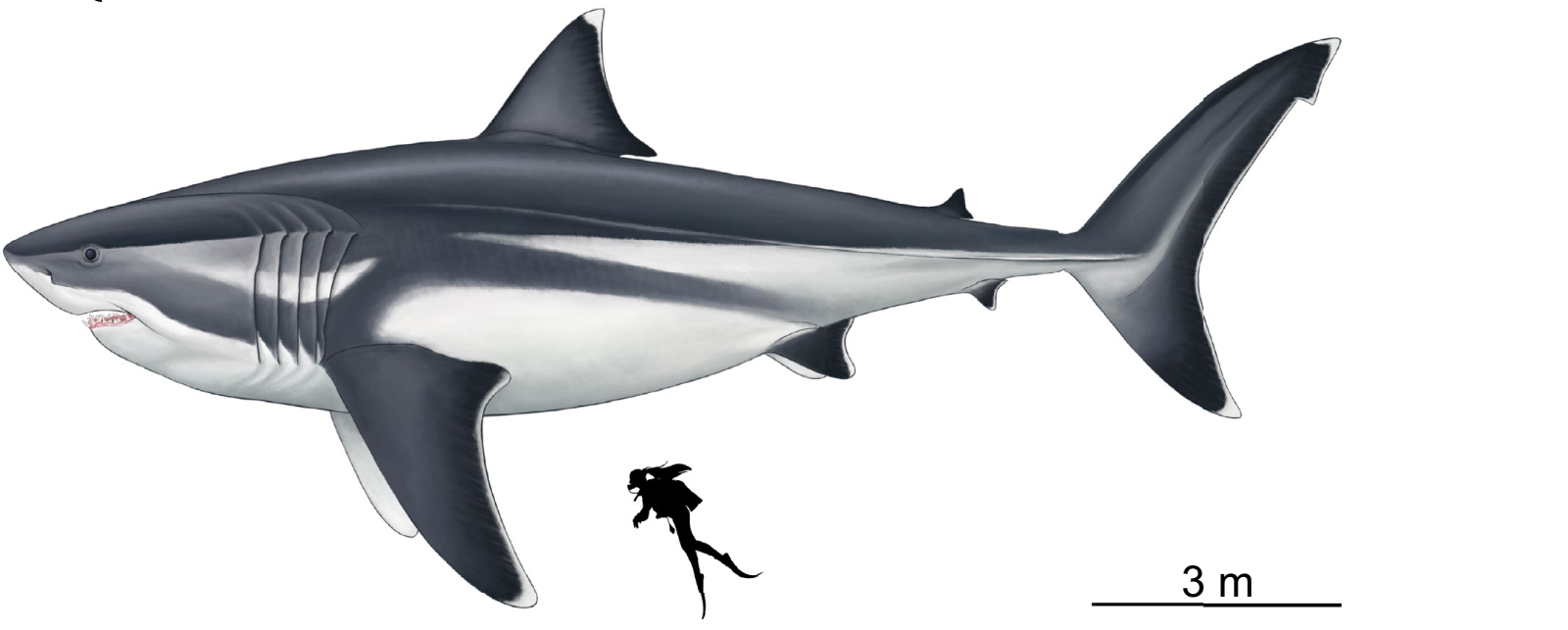}
			\caption{CAD construction, \citet{Cooper2020}}
			\label{Fig: Shark3}
		\end{center}
	\end{minipage}\newline
\end{figure}

CAD software enables designers to generate complex 3D models of animatronics, offering a comprehensive visual and structural plan for the final product. This facilitates the creation of elaborate patterns that are not only more realistic but also mechanically intricate, enabling the modelling of true movements. Several software applications can convert 2D sketches directly into 3D CAD models. Autodesk Fusion 360 allows you to import 2D sketches and use them to create 3D models, supporting sketch-based modeling on both Windows and Mac. SolidWorks offers powerful sketch tools for importing 2D sketches and converting them into 3D models using features like extrude, revolve, and sweep, primarily for Windows users. SketchUp is user-friendly, enabling quick conversion of 2D sketches to 3D models with import tools available for Windows, Mac, and as a web application. Rhino, known for its precision, imports various 2D file formats and converts them to 3D models, suitable for Windows and Mac. Tinkercad is a web-based tool ideal for beginners, allowing SVG file imports to create 3D models quickly and easily. Lastly, Shapr3D, designed for iPad and Mac, lets you sketch directly and convert those sketches into 3D models with an optimized touch interface.
CAD also facilitates the implementation of Prototyping. By leveraging rapid prototyping capabilities, designers can swiftly produce tangible objects from CAD designs utilizing precision manufacturing techniques such as CAM, which includes 3D printing or CNC machining. Testing and perfecting the mechanics of animatronics before full-scale production is essential \cite{Fei2012, Burns2016}. \citet{Anderson02} provide an initial exploration of the engineering and performance assessments of a robotic tuna, with a specific emphasis on its stability, maneuverability, and propulsion systems. It provides the requisite theoretical models, underscoring the benefits of fish-like propulsion for underwater vehicles. Moreover, CAD software allows for accurate digital reconstructions of the shark's anatomy using accessible data. \citet{Cooper2020} outlined the precise measurements of various shark body components, which can be entered into CAD software to produce very accurate three-dimensional models, see figures (\ref{Fig: Shark1}-\ref{Fig: Shark3}). These models can then be used to simulate motion and investigate biomechanics, which are essential for the development of animatronics that accurately replicate real-life motions.

\begin{figure}[tb]
	\noindent 
	\begin{minipage}[tb]{0.49\textwidth}
		\begin{center}
			\includegraphics[width=0.90\textwidth]
			{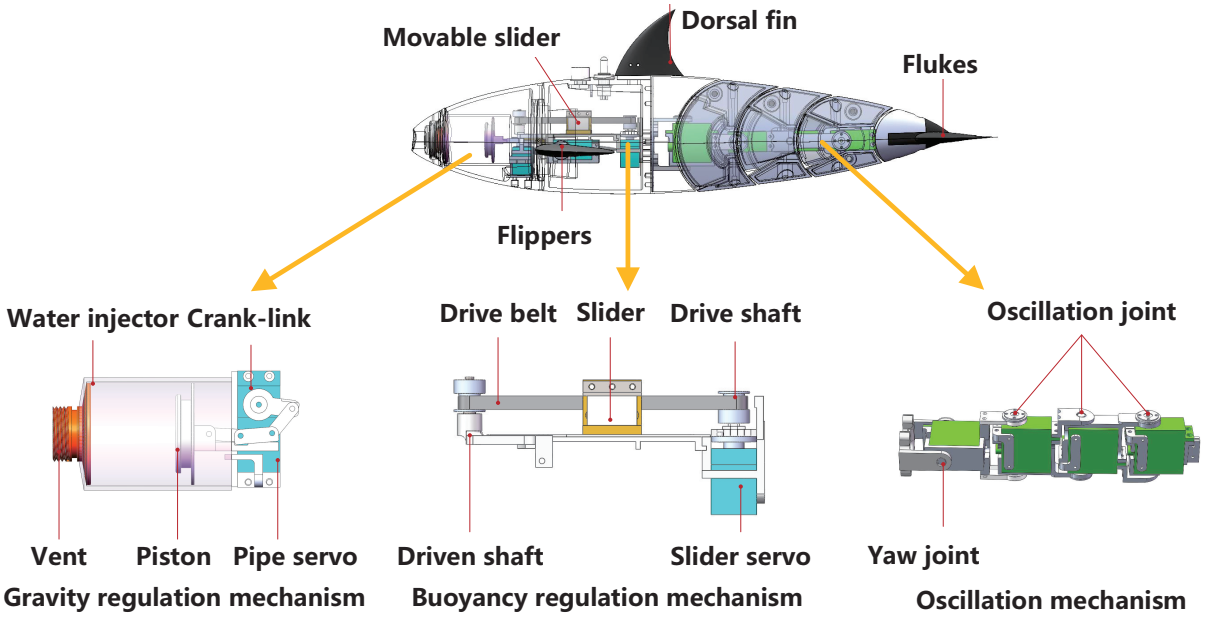}
			\caption{Bionic gliding robotic dolphin, \citet{Zhang2023}}
			\label{Fig: dolphin1}
		\end{center}
	\end{minipage}\noindent 
	\begin{minipage}[tb]{0.51\textwidth}
		\begin{center}
			\includegraphics[width=1.01\textwidth]
			{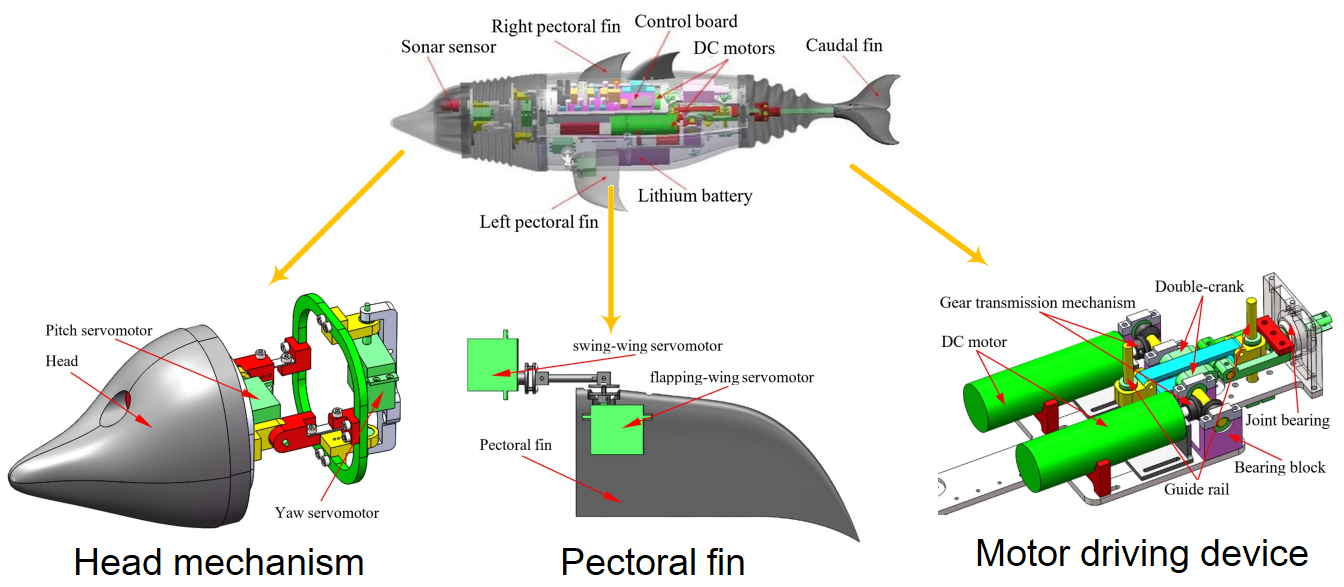}
			\caption{ElectroMechanical robotic dolphin, \citet{Yang2022}}
			\label{Fig: dolphin2}
		\end{center}
	\end{minipage}\newline
\end{figure}

Electromechanics, the intersection of mechanical engineering and electronics, impacts film-making, particularly through advanced animatronics. This area of study develops dynamic, interactive mechanical characters from multiple disciplines. Combining electrical control systems with mechanical hardware does this. Electromechanics uses sensors, actuators, and motors to simulate realistic movements and facial expressions in film animatronics \cite{Balakarthikeyan2023}. This interface allows for more complex animations and enhances animatronic figure reliability and efficacy. Electromechanical systems help film-makers manage sensitive face expressions and sophisticated body actions. This gives inanimate items lifelike and convincing properties, improving the viewer's experience. Electromechanics expands producers' artistic options and stretches cinematic special effects and character design \cite{YOKOI2011,Oka2014,Huber2021,Zhang2023,Chen2023,Chen2023B}. In this context, a research teams have succeeded in developing a robotic dolphin that mimics the complex movements of a real dolphin using advanced electromechanical systems \cite{Zhang2023, Yang2022}. This was accomplished by incorporating sensors and actuators, which play a vital role in generating realistic and lifelike movements, see figures (\ref{Fig: dolphin1}, \ref{Fig: dolphin2}). \citet{Wenqian2023} and \citet{Cui2023} provide a comprehensive survey of robotic fish, detailing various designs, propulsion mechanisms, and the advantages of undulating propulsion in harsh marine environments. Additionally, they emphasize theoretical, numerical, and experimental investigations, while offering future outlooks for underwater biomimetic robots.

\begin{figure}[tb]
	\noindent 
	\begin{minipage}[tb]{0.44\textwidth}
		\begin{center}
			\includegraphics[width=0.90\textwidth]
			{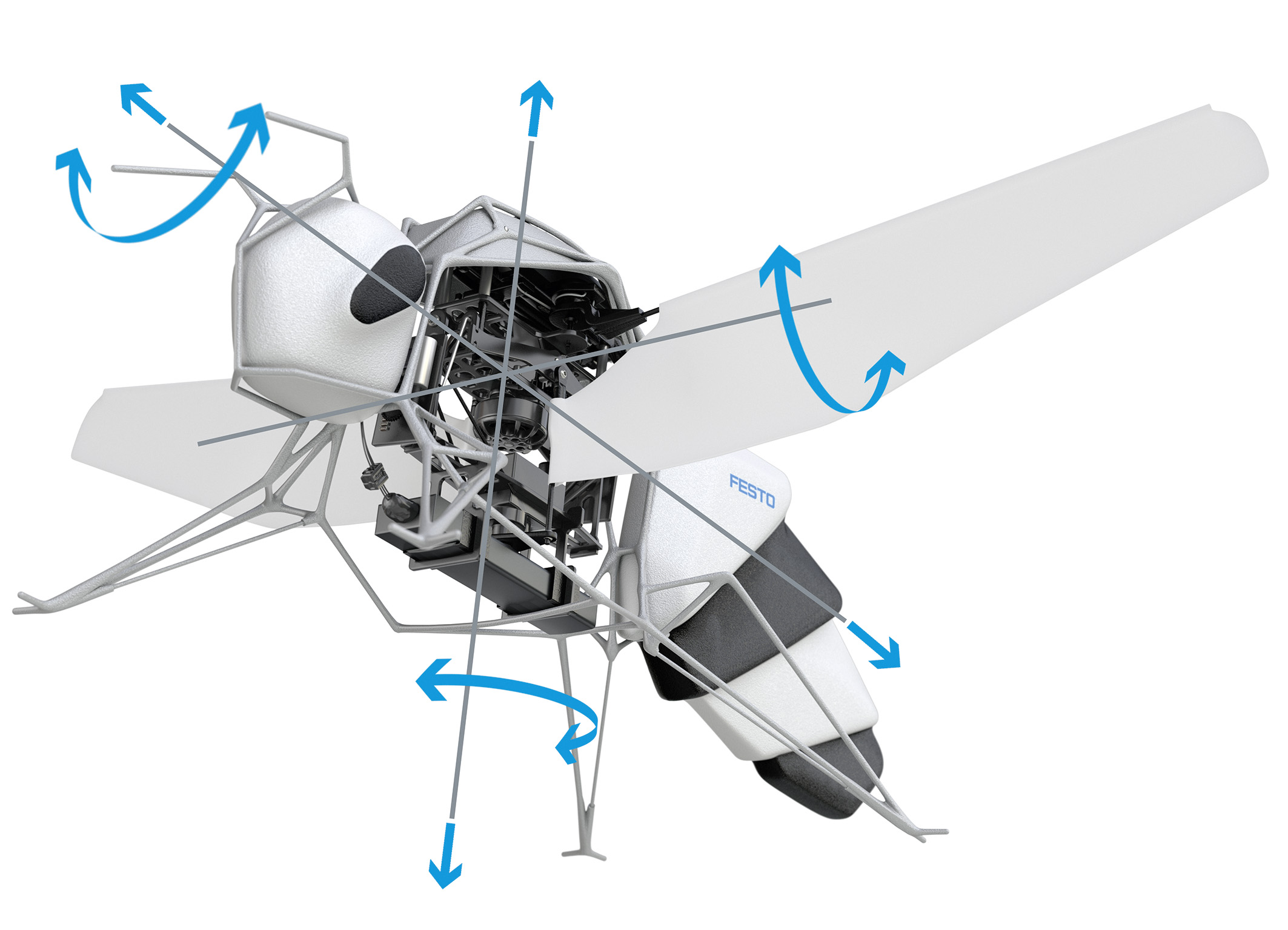}
			\caption{BionicBee, \citet{festo_bionicbee}}
			\label{Fig: Bee1}
		\end{center}
	\end{minipage}\noindent 
	\begin{minipage}[tb]{0.48\textwidth}
		\begin{center}
	\includegraphics[width=0.90\textwidth]
	{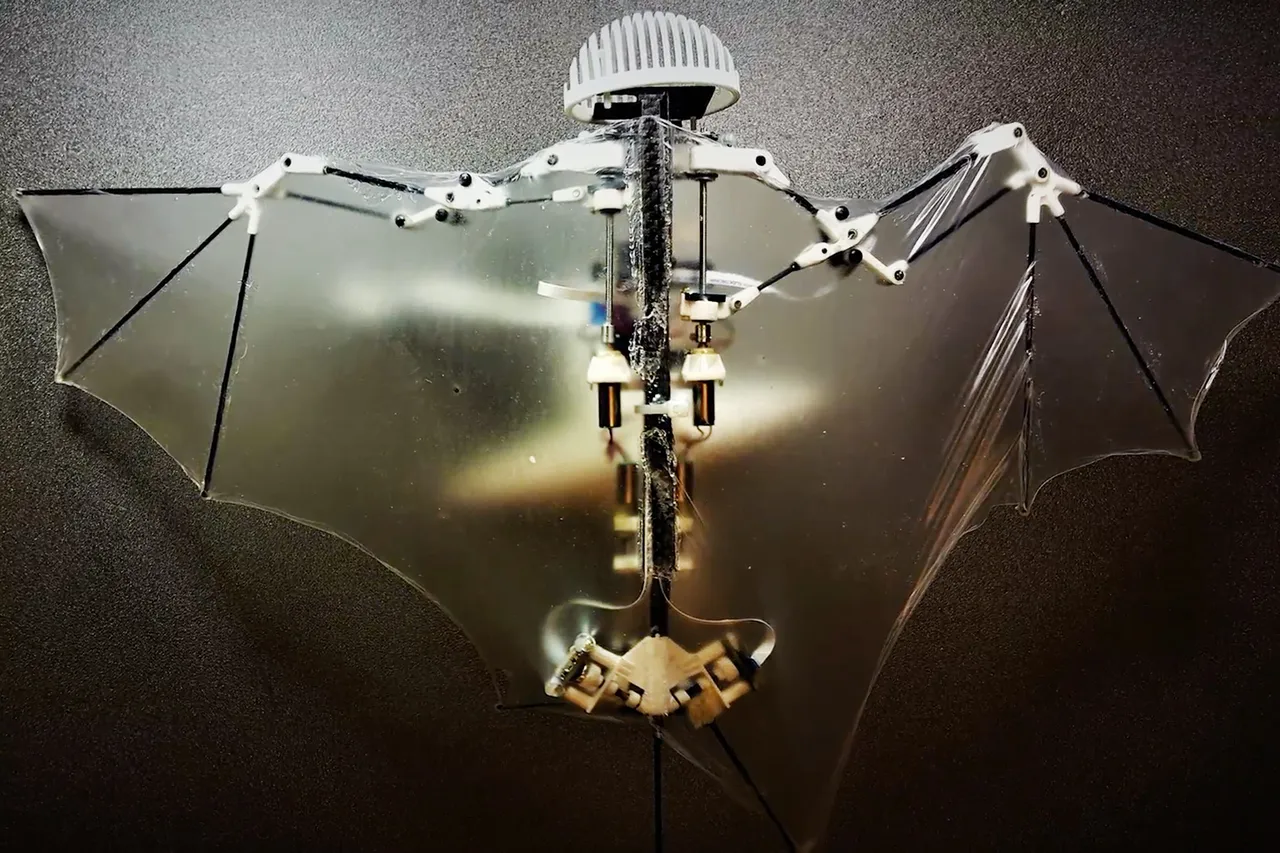}
	\caption{Robotic Bat, \citet{Ramezani2017}}
	\label{Fig: Bee2}
\end{center}
	\end{minipage}\newline
\end{figure}

Computational control methods extensively leverage Neural Networks (NN) and Reinforcement Learning (RL). Neural networks are instrumental in system identification, function approximation, and adaptive control, while reinforcement learning offers robust frameworks for optimizing control policies and decision-making processes \cite{Florian2018}. The integration of these technologies has led to significant advancements in the management of complex, dynamic systems, as evidenced by extensive research in the field. Additionally, swarm optimization algorithms, such as Particle Swarm Optimization (PSO) and Ant Colony Optimization (ACO), are utilized to fine-tune the hyperparameters of RL algorithms, thereby optimizing learning rates, exploration strategies, and network architectures \cite{Akhloufi2023,Arash2024}. In this context, Festo, a German company of automation technology and technical education, has developed the BionicBee as part of its Bionic Learning Network. The BionicBee represents an innovative approach to autonomous flying objects, designed to mimic the flight capabilities and behaviors of bees, see Fig.(\ref{Fig: Bee1}). This project embodies Festo’s commitment to learning from nature to inspire technological advancements. The BionicBee's wings beat at a frequency of 15 to 20 hertz, driven by a brushless motor. The wingbeat frequency can be adjusted to control lift and maneuverability. The three servo motors at the wing root alter the wing geometry, enhancing lift efficiency during different flight maneuvers. The bees achieve autonomous swarm flight using an indoor locating system based on ultra-wideband (UWB) technology. Eight UWB anchors installed at two levels provide accurate positioning by sending signals to the bees, which then calculate their positions using time stamps. A central computer specifies the flight paths, ensuring collision-free and coordinated swarm movements. Similarly, the kinematics and control of bat-like animatronics can be effectively modeled and optimized for realistic and efficient flight performance \cite{Ramezani2017,Colorado2018,DUAN2023}.

Control circuits and electronics regulate the flow of electrical signals, guaranteeing coordinated motions and responses. Microcontrollers and Microprocessors are essential components of animatronic systems, serving as the central processing units. They receive input from sensors and provide orders to actuators. Advanced microcontrollers provide intricate behavioural patterns and flexibility. Moreover, the integration of electronics with control systems in animatronics is a multidisciplinary endeavour that enhances the functionality and realism of cinematic effects. In Jurassic Park (1993), the film employed highly advanced animatronics to bring dinosaurs to life. The character, T-Rex, for instance, used complex control circuits to manage its massive movements and realistic appearances. Additionally, a combination of hydraulic systems and electronic control systems was used to synchronize movements and ensure realistic performances. Despite significant advancements, challenges remain in the miniaturization of components, power management, and the development of more intuitive control algorithms. Future research aims to address these challenges, further pushing the boundaries of what is possible in animatronic design \cite{whissel2010digital,JOCHUM2017}.

Therefore, the implementation of mechatronics, which means integrating mechanical components, electronics, control systems, CAD/CAM/CAE technologies, and computational controls, can introduce essential tools for creating highly functional and realistic animatronics. Each component plays a crucial role, and their seamless integration ensures that animatronics can perform complex, lifelike movements that enhance the viewer's experience. Future advancements in these fields will continue to push the boundaries of what is possible in animatronic design, leading to even more immersive and believable cinematic experiences.


\section{Animatronics in Movies}
Animatronics has played a pivotal role in enhancing the realism and emotional depth of cinematic experiences. This technology allows for the creation of lifelike robots and creatures that interact seamlessly with human actors, providing audiences with a more immersive and believable portrayal of fantastical elements. From the early days of cinema to the present, animatronics has evolved significantly, contributing to various genres including science fiction, horror, and fantasy. The table below showcases notable examples of animatronics in films, spanning from 1973 to 2024. It highlights the diverse applications of this technology across different models such as humans, sharks, bees, bats, monsters, and vampires. This comprehensive overview includes a total of 25 examples, demonstrating the extensive use and continual innovation of animatronics in the film industry over the past five decades.
\begin{longtable}{|>{\columncolor{lightblue}\centering\arraybackslash}m{2.0cm}|>{\raggedright\arraybackslash}m{11.0cm}|>{\centering\arraybackslash}m{2.0cm}|}
	\caption{Applications of Animatronics in Film-making} \\
	\hline
	\rowcolor{lightblue}
	\rule{0pt}{2.5ex} \textbf{Model} & \rule{0pt}{2.5ex} \textbf{Examples of Movies \& Description} & \rule{0pt}{2.5ex} \textbf{Reference} \\
	\hline
	\endfirsthead 
	
	\hline
	\rowcolor{lightblue}
	\rule{0pt}{2.5ex} \textbf{Model} & \rule{0pt}{2.5ex} \textbf{Examples of Movies \& Description} & \rule{0pt}{2.5ex} \textbf{Reference} \\
	\hline
	\endhead
	
	\hline
	\endfoot
	
	\hline
	\endlastfoot
	
	\multirow{1}{=}{\vspace{0.0cm} {Humanoids}} & 
	\textbf{Westworld (1973):} Science fiction film, which inspired the modern HBO series, American premium television network, featured animatronic robots in a futuristic amusement park. The film showcased advanced humanoid robots that interact with guests \cite{Cooper2020}. & \cite{Westworld} \\
	\cline{2-3}
	& \textbf{Alien (1979):} Combines the sculpture work of H.R. Giger and Carlo Rambaldi's mechanical design. The xenomorph's jaw and inner mouth at the tip of its tongue were mobile, including the latter's teeth. The creature's head is made of an impressive 900 moving parts. The biomechanical nightmare moves slowly and gracefully on top of its staggering seven-foot stature & \cite{Alien} \\
	\cline{2-3}
	& \textbf{Star Wars: Episode V - The Empire Strikes Back (1980)}: Walkers characters were built in various scales and in the picture they look as big as camels and to be animated  the armatures are actually built into the whole design these the joints the joints bend with constant friction.  &\cite{Star} \\
	\cline{2-3}
	& \textbf{E.T. the Extra-Terrestrial (1982)}: Bringing e.t to life required the work of animatronics.  inside a specially designed e. t suit and a professional mime with long fingers was hired to wear sleeve length e.t gloves to film shots that required intricate hand movements et's voice was provided by the actress pat welsh. &\cite{ET}\\
	\cline{2-3}
	\multirow{1}{=}{\vspace{0.0cm} {Humanoids}} & \textbf{The Thing (1982)} A Hand puppet built a really with an animatronic face between himself were the foam latex appendages for these insert shots these tentacles are small whips also operated by Boateng the goo being shot at the dog is carbacol a thickening agent. & \cite{Thing}\\
	\cline{2-3}
	& \textbf{The American Adventure (1982)}: Animatronics was used for the first time in a show in Disneyland called the Enchanted Tiki Room where birds sing and flowers croon as they say well the New York World's Fair was really a watershed in the development of audio animatronics. & \cite{American} \\
	\cline{2-3}
	
	& \textbf{The Terminator (1984)}: The other star of the Terminator a 200-pound electronically controlled robot created in the mind of director James Cameron built and designed by Stan Winston the robot stands 6 feet 2 inches tall and has over a thousand chrome-plated moving parts that were handcrafted by Stan and his staff building this mechanical marvel took almost a year from the original idea and initial drawings Stan and his crew prepare for a scene where the cyborg has been injured in a fiery explosion. & \cite{Terminator} \\
	\cline{2-3}
	& \textbf{The Adventures of Teddy Ruxpin ( 1985 )}: Teddy Ruxpin was the world's first animatronics storytelling toy and one of the toys of childhood past , so they decided to call him an Illya this animatronic little stinker had the ability to blink and move his mouth . Teddy Ruxpin who rose to save the company 1987 brought baby Teddy Ruxpin an on cassette plane eye and mouth moving animatronic that would respond to noises in the environment with phrases the answer box was a contraption with four buttons that came with a booklet and Teddy would ask children to answer questions corresponding to the pictures in the booklet with their button answers and of course there was the Teddy Ruxpin picture show a slide projector that paired with Ruxpin. & \cite{Adventures} \\
	\cline{2-3}
	& \textbf{Aliens (1986)}: The Queen alien was that was a new really huge creature big experience for us so i don't think i can't remember before that was done not not a whole lot of animatronic effects have been done to that scale up to that time maybe dragon slayer had some elements elements that were close but not to that high action of of 20,000 leagues on the seas yeah big hydraulic tentacles most of the work that we did with the queen was with the one hero puppet. & \cite{Aliens}\\
	\cline{2-3}
	& \textbf{Predator (1987)}:  The original Predator's mechanical face was brought to life with servo-driven cable-actuated mandibles and eyebrows at Stan Winston Studio. It was a three-jointed mandible, so all the tips moved, and they would swing out, but the bottom didn't spread huge, massive" Tonegawa-Seiko SSPS-102 servo motor&\cite{Predator}\\
	\cline{2-3}
	& \textbf{Men in Black (1997)}: The oversized worm guy from Barry Sonnenfeld sci-fi classic men in black 2 now this piece was made oversized so that it can be featured in the close-up shot it is movable imposable for the time being the foam latex on this thing is very soft still the puppet is supported by two metal poles that are connected to a wooden base it has hand control for the arms and the digits and those are controlled back here now this is a very cool piece because it's three times the size of the other puppets that were used in the wide shots of the film this features a ton of detail in the face and the hands and it's movable .  &\cite{Men}\\
		\cline{2-3}
	& \textbf{Hulk (2003)}: A full animatronic was never created as it was decided it would complicate production to set up shots for a puppet and then a computer graphic. An animatronic was used for Sterns' mutating head. &\cite{Hulk}\\
		\cline{2-3}
	\multirow{1}{=}{\vspace{0.0cm} {Humanoids}} &  \textbf{Hellboy (2004)}: The Character Samuel's tentacle was made by animatronics designer Mark Setrakian. In Hellboy (2004), he recognized that the cascading quality could be adapted to the scrunt's tail. When the delay is set to zero, the tail coils up, increases the delay, and wags like a genuine tail. The cascade controller distributes motion from finger to finger. Mark constructed the illusion of a finger drum on King Balor's hand. &\cite{Hellboy}\\
		\cline{2-3}
	& \textbf{Hellboy II: The Golden Army (2008 )}: The cascade controller also works on fingers. By distributing motion from finger to finger, Mark – the roboticist - created the illusion of finger drumming on King Balor's hand.  &\cite{Hellboy2}\\
		\cline{2-3}
	& \textbf{Hansel and Gretel: Witch Hunters (2013)}: The performer who played Edward the troll wore two animatronic hands, but they were so large that the performer's hands only reached the troll's wrists. As he moved the hands around, the subtle finger motions were created with radio-controlled hobby servos, like you'd find in a model plane. Each hand had 10 servos spread throughout the fingers.&\cite{Hansel}\\
		\cline{2-3}
	& \textbf{The Hug (2018)}: An animatronic panda at a rundown pizza place seems to be out of order - until little Aden gives it a hug. &\cite{imdbShort2018}\\
		\cline{2-3}
	\hline
	\multirow{1}{=}{\centering {Puppets}} & 
	\textbf{Child's Play (1988)}: It features the work of puppet designer Kevin Yagher. The Good Guy doll with a serial killer's soul was operated by nine puppeteers and radio-controlled animatronics. The doll's movements begin as innocent and cherub-like, then quickly evolve into more human-esque movements. Different Chucky dolls were made to fulfill physical feats as well. \vspace{0.25cm} & \cite{Child}\\
	\cline{2-3}
	& \textbf{Ghostbusters (1989)}: Prop studio, The Character Shop, headed by Rick Lazzarini, who had worked on Ghostbusters II, was hired to take on the task, with their incarnation of Slimer featuring a highly detailed design, along with animatronics that controlled the key features found on the face, and his tongue. & \cite{Ghostbusters} \\
	\cline{2-3}
	& \textbf{Willy's Wonderland (2021)} : The Animatronics are the main antagonists of the 2021 horror comedy film Willy's Wonderland. They were once friendly animatronics and mascots at Willy's Wonderland, owned by serial killer Jerry Robert Willis, and seemingly designed by local mechanic Jed Love who united equally sadistic killers to assist him in murdering families within the establishment. To avoid arrest, the killers perform a satanic suicide ritual to transfer their souls into the animatronics, possessing them and actively killing and devouring everyone in their sight.
		 & \cite{Willy} \\
	\cline{2-3}
	& \textbf{Five Nights at Freddy's (2023)}: American supernatural horror film follows a troubled young man caring for his 10-year-old sister who is suggested by his career counselor to take up a night shift job at a once-successful abandoned family entertainment center, while keeping an eye on the restaurant's four murderous animatronic characters Freddy Fazbear, Bonnie, Chica, and Foxy (possessed by the spirits of five missing dead children) while trying to figure out unsolved disappearance of his younger brother more than a decade before. It is based on the Five Nights at Freddy's video game series created by Scott Cawthon& \cite{Five} \\
			\cline{2-3}
	\hline
	\multirow{1}{=}{\centering {Sharks}} & 
	\textbf{Jaws (1975):} Bruce the shark in Jaws is a combination of three animatronic sharks: one that moved from left-to-right; another which moved from right-to-left, and a shark sled. The shark sled was an animatronic shark attached to a large arm that went back and forth along an underwater rail mounted to the seafloor. The crew moved this rig to various locations off the coast of Martha’s Vineyard to film the different scenes in the script. & \cite{Jaws} \\
	\cline{2-3}
	\multirow{1}{=}{\centering {Sharks}}& \textbf{Jaws The Revenge (1987):} The sharks launched from atop an 88-foot long platform made from the truss turret of a 30-foot crane and floated out into clifton bay in addition to this seven sharks or segments were produced two models were fully articulated two were made for jumping one for ramming one was a half shark the top half and one just a fin the two fully articulated models each had 22 sectioned ribs and movable jaws covered by a flexible water-based latex skin & \cite{Jaws2} \\
	\cline{2-3}
	& \textbf{Deep Blue Sea (1999)}: An animatronic shark was developed utilizing aviation technology, and these sharks truly move underwater completely remote controlled on their own power, moving their tails, gills, eyes, jaw, and so on. & \cite{Deep} \\
	\hline
	\multirow{3}{=}{\centering {Animals}} & 
	\textbf{An American Werewolf in London (1981)} : the forehead and cheek sections of the piece move outwards and the jaw section extended forwards, opening the mouth. These actions would have been operated on set by a team of effects technicians using air-filled syringes. This mechanism has been sympathetically replicated using electric motors and cables, allowing the piece to cycle through a series of movements that recreate the motion seen in the film. & \cite{American2} \\
	\cline{2-3}
	& \textbf{Gremlins (1984)}: The small, furry creatures mutate into violent and destructive reptilian monsters when they are given water and fed after midnight. In the film's folklore, Being so small, the mechanics were more prone to malfunction. & \cite{Gremlins} \\
	\cline{2-3}
	& \textbf{Jurassic Park (1993):} Made dinosaurs roam the Earth again. The life-sized animatronics were created by special effects guru Stan Winston. Paleontologist Jack Horner oversaw the designs to make sure they were portrayed as animals, not monsters. The first dinosaur created for the franchise was a Tyrannosaurus Rex, affectionately named Rexy, which was 20 feet tall and 40 feet long and weighed 17,500 pounds. & \cite{Jurassic}\\
	\cline{2-3}
	& \textbf{Animal Farm (1999)}: An incredible accurately detailed full scale animatronic Pig Head, made for the 1999 film version of the classic George Orwell Novel. & \cite{ANIMAL}\\
	\cline{2-3}
& \textbf{Lady in the Water (2006)}: Here are the Scrunt's hard mechanics. This piece here both pulls the nose back and lifts it up at the same time. But those hard mechanics had to work hand in hand with the softer elements of the wolf's face, like these lips that pull up and expose the fangs.& \cite{Lady}\\
\cline{2-3}
& \textbf{The Banana Splits Movie (2019)}: Based on the 1968-1970 children's television series of the same name. These robotic mascots are created by Karl and go on a killing spree after learning about the show's cancellation.& \cite{BananaSplitsMovie}\\
\cline{2-3}
& \textbf{Olaf, Frozen Ever After (2016)}: he likes warm hugs, and he’s one of the most technologically advanced Audio-Animatronics® figures at Walt Disney World® Resort! Guests may feel like Olaf walked right out of the film Frozen and into the attraction with his fluid movements and facial expressions as he sings to guests.& \cite{OlafFrozenEverAfter}\\
\hline	
	\multirow{1}{=}{\centering {Birds}} & 
	\textbf{Mary Poppins (1964)}: is the Disney musical fantasy about a dysfunctional London family that employs the titular nanny to help sort out their lives. the animatronic bird keeps the magic grounded in reality.& \cite{MaryPoppins}\\
	\hline
        \multirow{1}{=}{\centering {Creatures}} & 
	\textbf{King Kong (1976)}: The giant ape (both a man in a suit and a 40-foot animatronic) is enamored by her beauty as he extends his mechanical arms and legs. & \cite{KingKong}\\
	\cline{2-3}
	& \textbf{Labyrinth (1986)}: The puppets were created by Jim Henson's Creature Shop, one of which befriends Sarah, an ogre named Ludo. Weighing 75 pounds, Ludo's performance was shared by artists Ron Mueck and Rob Mills. & \cite{Labyrinth} \\
	\cline{2-3}
	
	\multirow{1}{=}{\centering {Creatures}} & \textbf{Hellboy (2004)}: A very tricky motion challenge has been encountered : cascading. These complex motions are used for movements like wagging or the curling of a tentacle. So Mark linked up servos in a chain, then created a delay so each servo moved a little bit after the one before it, causing a chain reaction down the piece. The more servos, the smoother the motion & \cite{Hellboy3}\\
	\cline{2-3}
	& \textbf{Stranger Things ( 2016-2025 )}: These are cables embedded in the outer layer of animatronics. Soft mechanics are often used for animatronics that don't have traditional bones and joints, like the petals on the Demogorgon's head  & \cite{StrangerThings}\\
	\cline{2-3}
	& \textbf{The Tomorrow War (2021)}: A terrifying White Spike animatronic that the Ropotist helped create can tilt its head, thanks to major technological advancements. This six-axis module has six degrees of freedom, t can turn, it can tilt, and it can also shift, and it can change length, too, which is very unusual. the White Spike is moved by joysticks.&\cite{TomorrowWar} \\
	\cline{2-3}
	& \textbf{Ghostbusters ( 1984-2021 )}: In this shot from "Ghostbusters: Afterlife," a Terror Dog stomps on Paul Rudd's car. But that's not CGI. It's actually a puppeteer wearing a mechanical Terror Dog arm controlled by a series of cables. To make creatures like the Terror Dog and this ghoulish skeleton hyperrealistic, makeup effects teams have to juggle visual design, classic puppeteering, and animatronics. & \cite{Ghostbusters2}\\
	\hline
\end{longtable}

\section{Exploring the Influence of Mechatronics on Creating Lifelike Movie Characters}
In the development of animatronic creatures for films, such as \textit{The Tomorrow War} (2021), \textit{Hansel \& Gretel: Witch Hunters} (2013), and \textit{Hellboy} (2004), significant attention is given to the realism of movement to enhance the believability of these characters. The integration of servos, soft mechanics, and advanced motion control systems plays a critical role in mimicking natural movements, such as finger drumming and tail wagging, which are essential for conveying lifelike behavior. The collaboration between mechanical design and artistic sculpting is crucial, with precise anatomical considerations ensuring that each muscle and joint functions accurately within the animatronic framework. Recent technological advancements, particularly in digital industrial servos, have further enabled more complex and precise motions, allowing for a higher degree of realism, as demonstrated in the White Spike animatronic, which exhibits sophisticated head-tilt and other dynamic actions. This interdisciplinary approach underscores the importance of combining engineering expertise with an understanding of biological motion to create convincing animatronic characters in cinematic applications. This is apparent in the animatronic shark technology used in six different movies: \textit{Jaws} (1975), \textit{Jaws 2} (1978), \textit{Jaws 3} (1983), \textit{Jaws: The Revenge} (1987), \textit{Deep Blue Sea} (1999), and \textit{The Meg} (2018). Below is a description of the differences between them:

\begin{figure}[!p!b]
	\noindent 
	\begin{center}
		\begin{minipage}[tb]{0.48\textwidth}
			\includegraphics[width=0.90\textwidth]
			{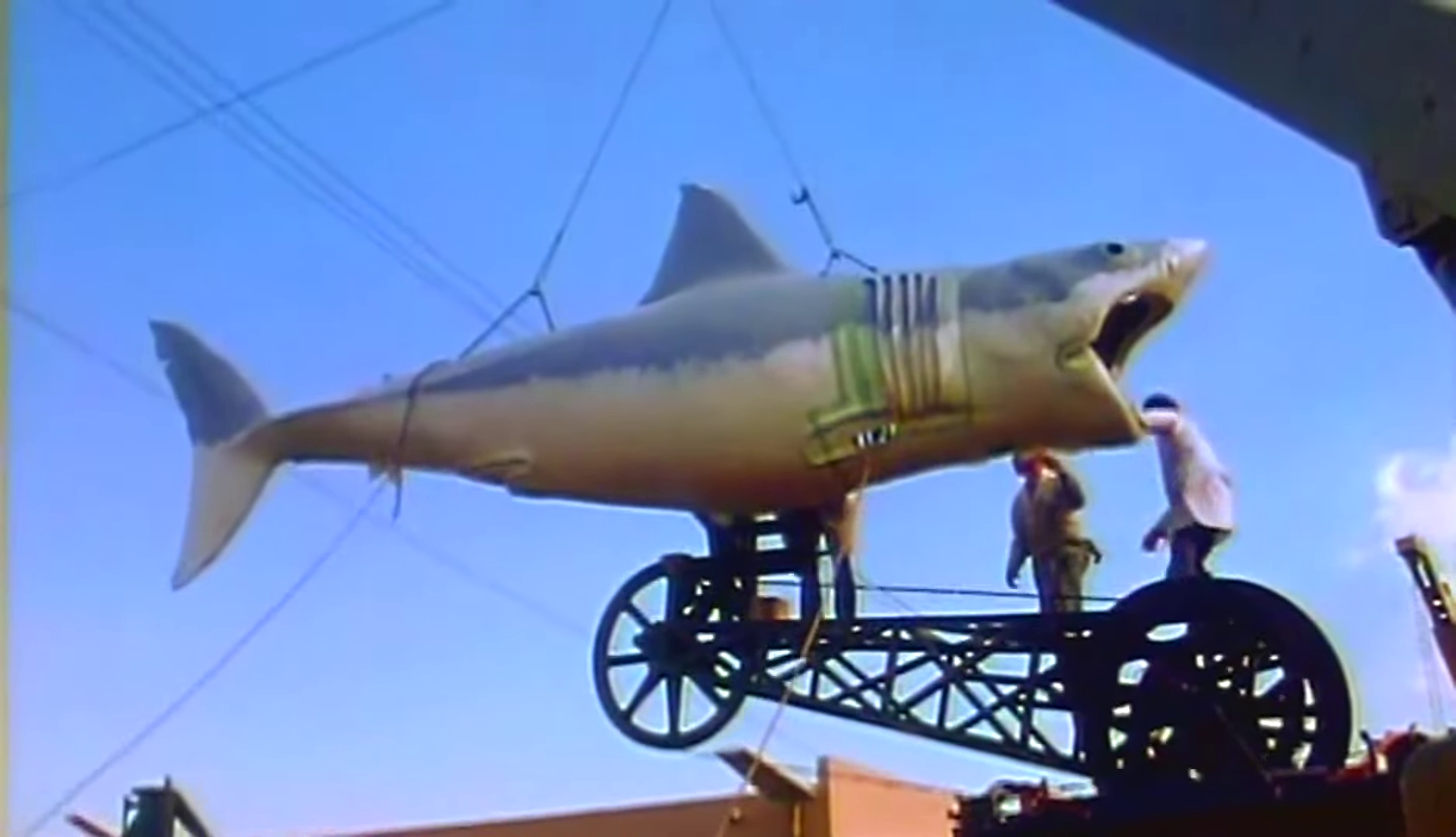}
			\caption{Jaws 1975: Mechanical sharks}
			\label{Fig: Jaws1975_Mech}
		\end{minipage}\noindent 
		\begin{minipage}[tb]{0.46\textwidth}
			\includegraphics[width=0.85\textwidth]
			{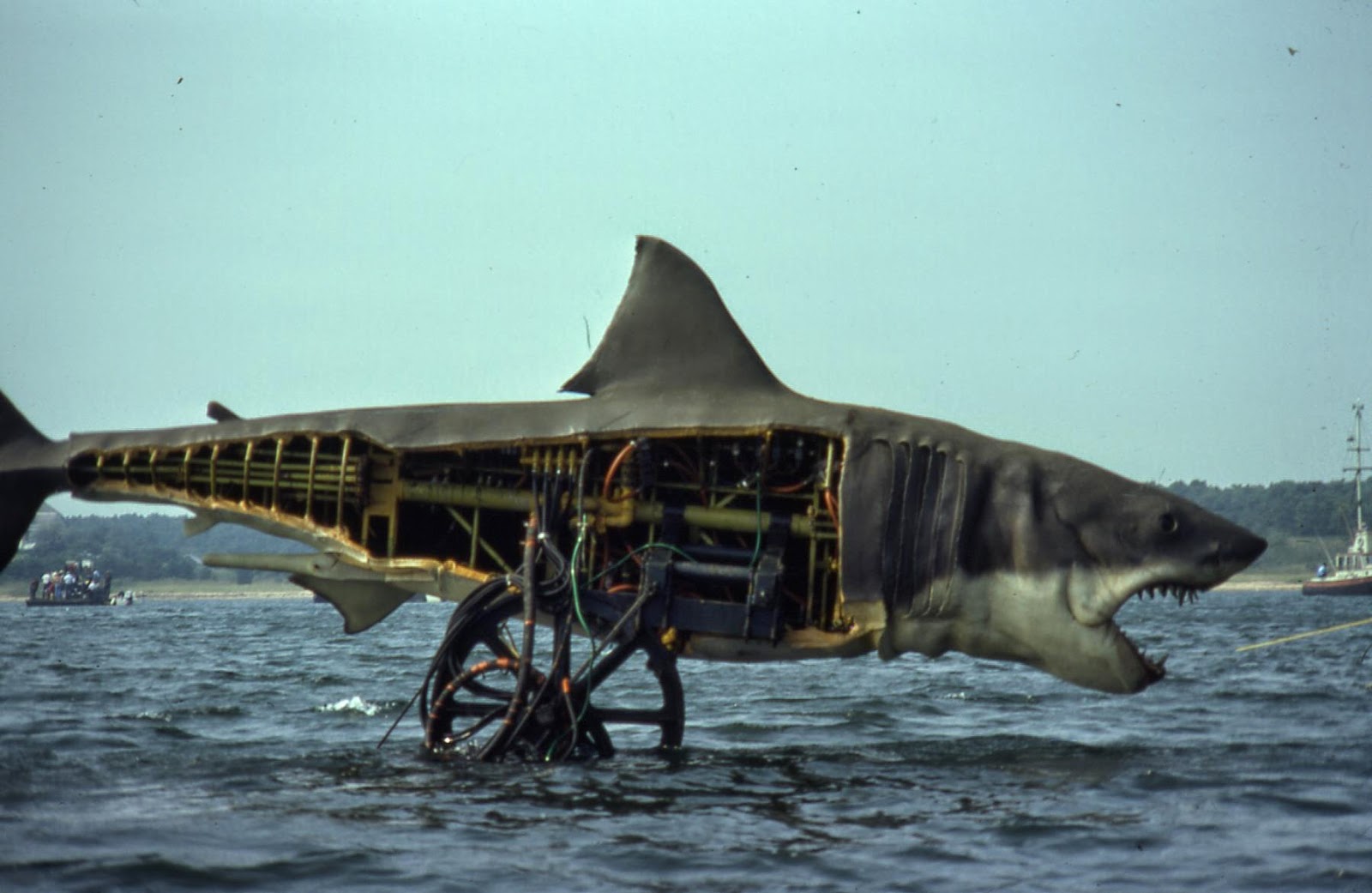}
			\caption{Bruce, the mechanical shark}
			\label{Fig: Jaws1975_Mech2}
		\end{minipage}
	\end{center}
\end{figure}

\begin{figure}[tbh]
	\noindent 
	\begin{center}
		\begin{minipage}[tb]{0.46\textwidth}
			\includegraphics[width=0.90\textwidth]
			{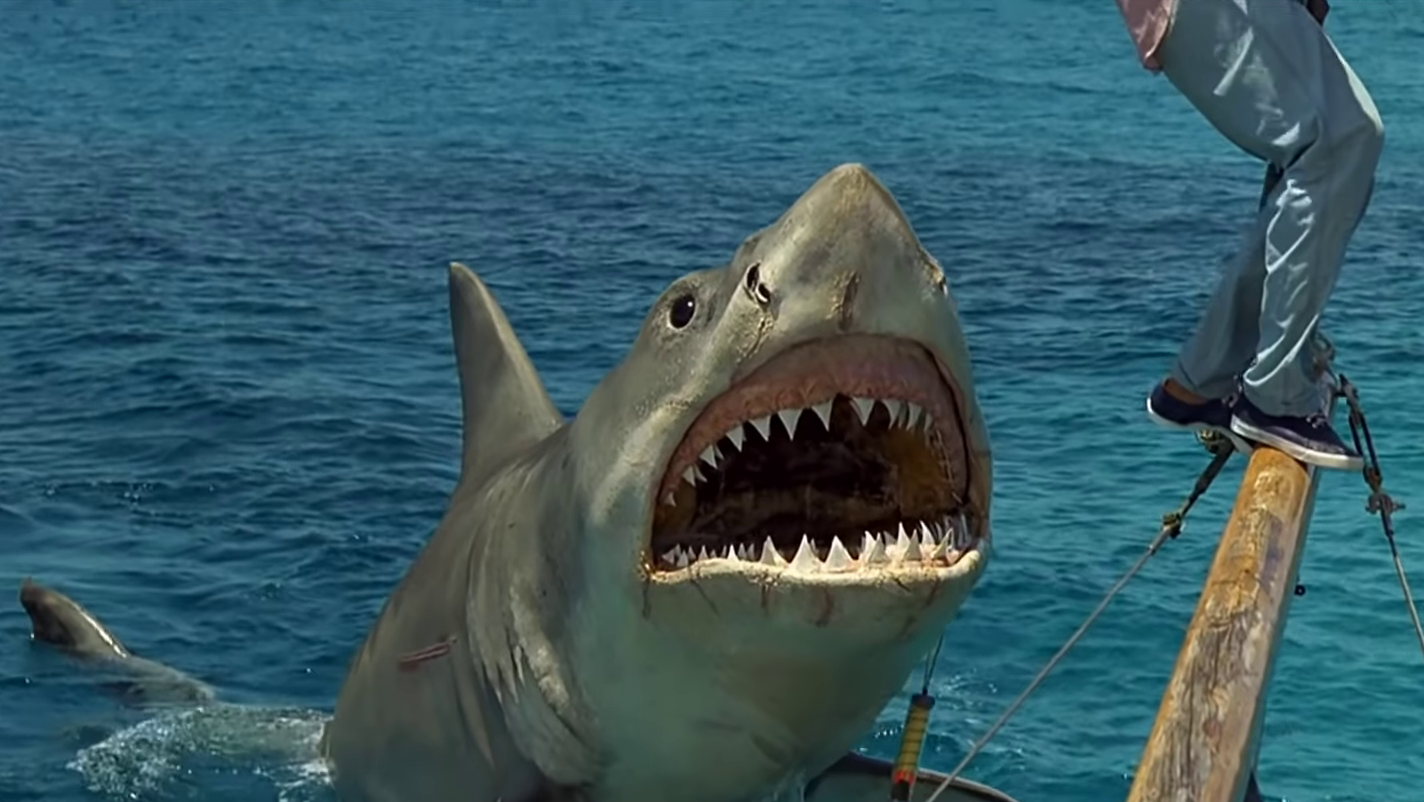}
			\caption{Jaws: The Revenge (1987)}
			\label{Fig: Jaws1987}
		\end{minipage}\noindent 
		\begin{minipage}[tb]{0.44\textwidth}
			\includegraphics[width=0.80\textwidth]
			{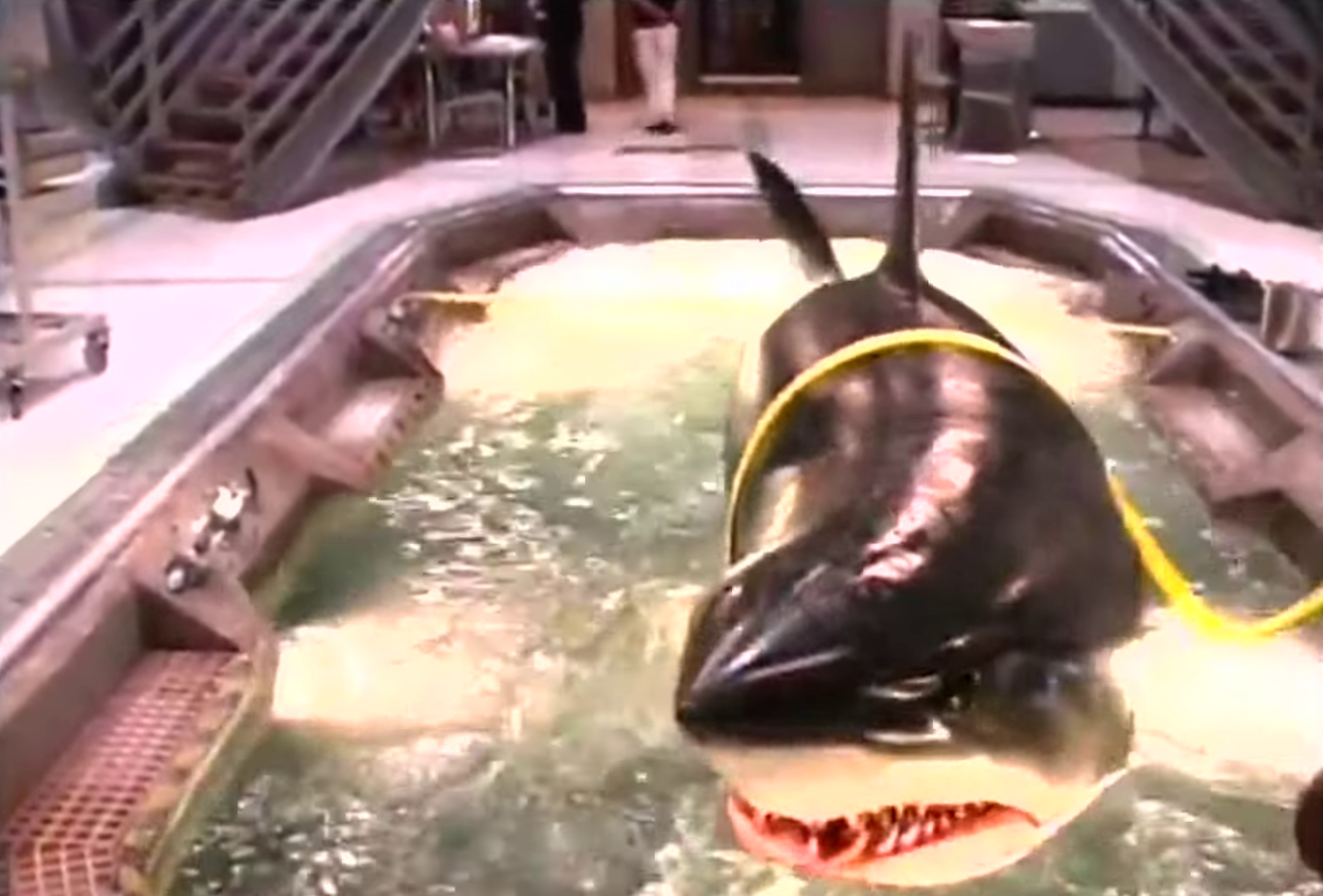}
			\caption{Deep Blue Sea (1999)}
			\label{Fig: Deep_Blue_Sea}
		\end{minipage}
	\end{center}
\end{figure}

\begin{itemize}
	\item \textit{Jaws (1975)}: The model used air-powered pneumatic mechanisms for the 25-foot animatronics, which weighed hundreds of pounds. The mechanical sharks were constructed with steel skeletons covered by polyurethane rubber skin, see Fig.(\ref{Fig: Jaws1975_Mech}). The movement was provided by air-powered pneumatic mechanisms that worked absolutely perfectly, see Fig.(\ref{Fig: Jaws1975_Mech2}) ,without a single issue on land. The shark had issues with slipping off the platform and getting tangled in seaweed. The pneumatic hoses controlling the shark's movements corroded, and the salt water of the ocean caused flooding in the tubes, leading to electrical component failures and making the animatronics difficult to work with. These were animatronics that cost half a million dollars to produce, and most of the time, they just did not work, and one of them sank. In Jaws (1975), the iconic opening scene where a swimmer is attacked famously avoids showing the shark itself. This was a creative decision born out of necessity due to the difficulties with the animatronics. The film effectively described the attack without revealing the shark, resulting in a more convincing and terrifying portrayal. The most frightening aspect of the shark in Jaws is not the animatronics or the footage of actual great whites, but rather the fear of the shark that remains unseen.
	\item \textit{Jaws 2 (1978)}: Featured two sharks and an animatronic fin. The metal equipment's shark parts were damaged due to wind changes, indicating environmental challenges in controlling the animatronics.
	\item \textit{Jaws 3 (1983)}: Introduced a combination of miniatures and animatronic sharks mixed with CGI instead of using full-bodied animatronics. One shark was built as a cable-articulated puppet with some functions, but static motions were noted due to the lack of puppeteers.
	\item \textit{Jaws: The Revenge (1987)}: the sharks were primarily created using animatronics, designed and constructed to interact with the environment and actors in real-time. These mechanical sharks were operated using a combination of hydraulic and pneumatic systems, similar to the animatronics used in the earlier Jaws films. The production relied heavily on practical effects, including full-scale animatronic sharks, to create the appearance of the shark on screen, see Fig. (\ref{Fig: Jaws1987}). The animatronic sharks were launched from atop an 88-foot-long platform with a truss turret of a 30-foot crane. Each model had 22 sectioned ribs and movable jaws, indicating an advancement in mechanical complexity compared to earlier films.
	\item \textit{Deep Blue Sea (1999)}: The film utilized a combination of live sharks, highly advanced animatronic models, and CGI to bring the sharks to life. A combination of animatronics and Computer-Generated Imagery (CGI) was used to create the sharks. Featured an animatronic shark built with airplane technology, measuring 25 feet long and weighing 8,000 pounds. The sharks were completely remote-controlled, with their own power moving hydraulics and electronics, technology borrowed from aerospace to create realistic shark movements and behaviors, see Fig. (\ref{Fig: Deep_Blue_Sea}). This setup allowed for smoother movements and more advanced control compared to earlier films. CGI was used extensively in Deep Blue Sea for scenes where the sharks needed to perform actions that were too complex, dangerous, or impossible for the animatronic models. This included fast swimming sequences, jumping out of the water, and certain attack scenes. The CGI sharks were created using computer models and animation techniques, allowing for more dynamic and flexible movements in these action-heavy scenes.
\end{itemize}

\begin{figure}[tb]
		\begin{center}
	\begin{minipage}[tb]{0.75\textwidth}
			\includegraphics[width=0.90\textwidth]
			{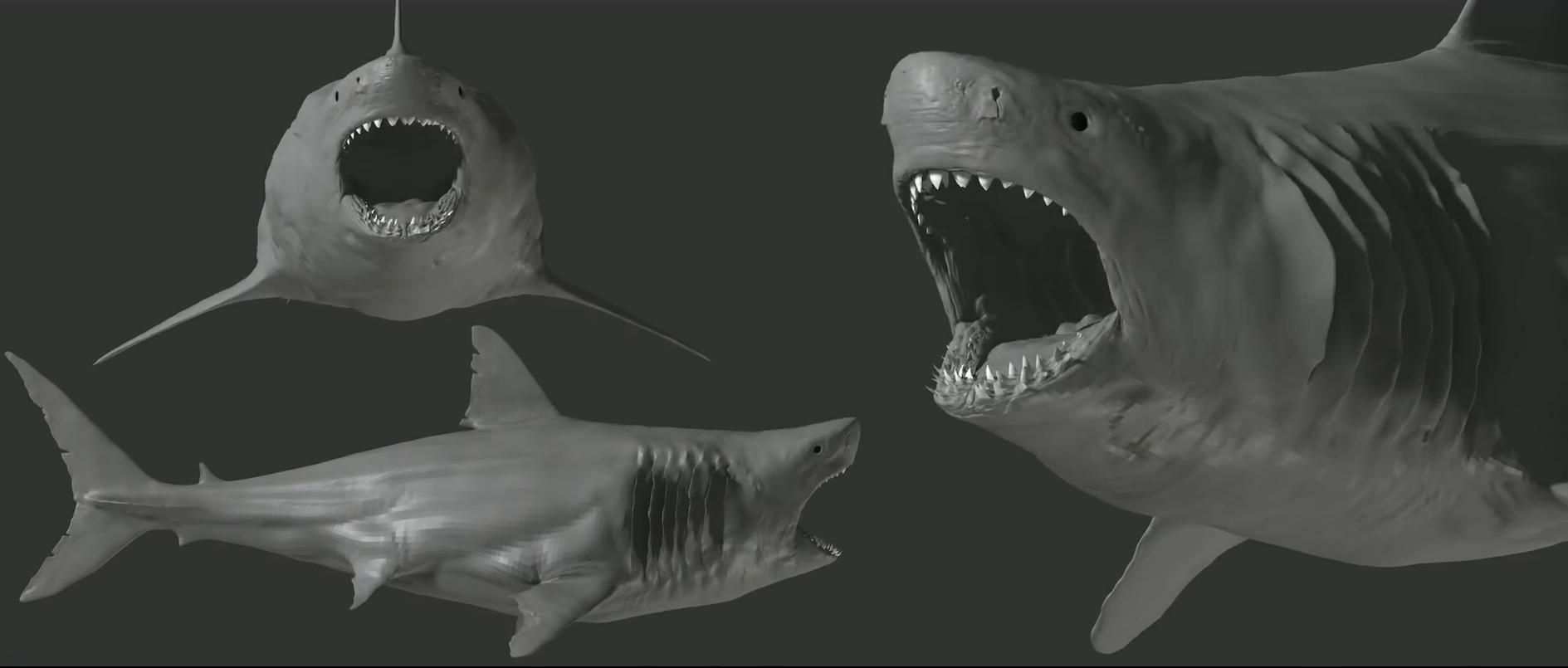}
			\caption{The Meg 2018: Computer-Generated Imagery of Meglodon}
			\label{Fig: TheMeg3}
	\end{minipage}
		\end{center}
\end{figure}

\begin{figure}[tb]
	\begin{center}
		\begin{minipage}[tb]{0.75\textwidth}
			\includegraphics[width=0.90\textwidth]
			{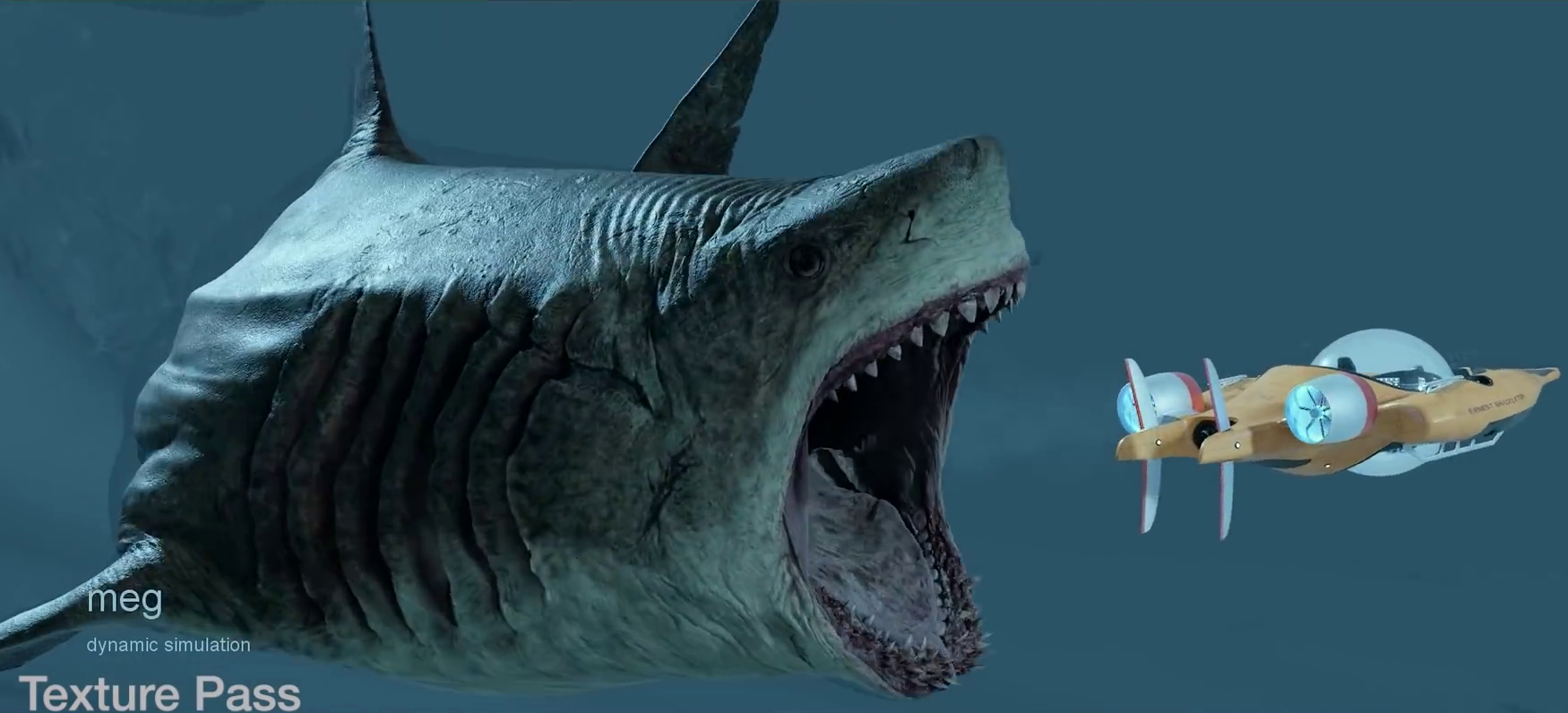}
			\caption{The Meg 2018: enhanced realism in portraying the massive shark}
			\label{Fig: TheMeg4}
		\end{minipage}
	\end{center}
\end{figure}

The primary differences across these films include advancements in animatronic technology, transitioning from basic pneumatic systems in \textit{Jaws} to more sophisticated remote-controlled mechanisms in \textit{Deep Blue Sea}. In contrast, \textit{The Meg (2018)} presented the "Megalodon," a prehistoric shark significantly larger than other sharks and marine animals, measuring 75 feet (23 meters) in length. Due to the enormous size and complexity of the Megalodon, the filmmakers opted to primarily utilize CGI rather than traditional animatronic models, see Fig.(\ref{Fig: TheMeg3}). The production team employed CAD models to meticulously design the shark, ensuring accuracy and intricate detail. These digital models were then animated with CGI to realistically interact with actors and environments. This approach allowed for greater flexibility and enhanced realism in portraying the massive, prehistoric shark during various action sequences, see Fig.(\ref{Fig: TheMeg4}).

The comparison between \textit{Jaws (1975)} and \textit{The Meg (2018)} highlights the advancements in technology over the decades, from the rudimentary, albeit groundbreaking, animatronics of \textit{Jaws} to the sophisticated, highly realistic creations in \textit{The Meg}. These developments have not only enhanced the visual experience for audiences but have also streamlined production processes, allowing filmmakers to bring their visions to life with greater precision and reliability.
On the other side, while the animatronics in \textit{The Meg (2018)} were technically superior, some critics argued that the film lacked the suspense and emotional impact of JAWS. The heavy reliance on CGI, although visually impressive, sometimes made the shark feel less tangible and real compared to the purely mechanical "Bruce" from \textit{Jaws (1975)}.

\section{State of the Art in Mechatronics for Modern Animatronics}
The latest advancements in mechatronic technologies, particularly the integration of flexible multibody dynamics through Absolute Nodal Coordinate Formulation (ANCF), have significantly enhanced the capabilities of animatronic creatures in film-making. These technological innovations continue to push the boundaries of what is possible in animatronic design, leading to more realistic and functional animatronic systems \cite{tesar2008,rolf2011}.

\subsection{Cutting-Edge Mechatronics for Film-making}
In recent years, mechatronics has seen significant advancements that have profoundly impacted the field of film-making, particularly in the creation and operation of animatronic creatures. These technological innovations span various domains, including material science, electronics, control systems, and computational methods, each contributing to more sophisticated and realistic animatronics. The development of flexible multibody dynamics, specifically the ANCF, has introduced new methods for accurately simulating and controlling the complex movements of animatronic components. This approach allows for the precise modeling of large deformations and rotations \cite{Nachbagauer2014, shabana2023state}, enhancing the overall realism and functionality of animatronics.  These recent aspects have positive impact on movements through various means, such as:

\begin{itemize}
	\item \textbf{Advanced Materials and Manufacturing Techniques}
	\begin{itemize}
		\item The development of new materials, such as flexible polymers and lightweight composites, has enhanced the durability and flexibility of animatronic components. These materials allow for more lifelike skin textures and more fluid movements \cite{seitz2005three}.
		\item Innovations in additive manufacturing, particularly 3D printing, have revolutionized the prototyping and production of animatronic parts \cite{lipson2013fabricated}. Rapid prototyping capabilities enable filmmakers to quickly iterate and refine designs, ensuring optimal performance and realism.
	\end{itemize}
	\item \textbf{Enhanced Electronics and Actuation Systems}
	\begin{itemize}
		\item Modern animatronics leverage advanced electronic components, including high-precision sensors and actuators, to achieve more accurate and responsive movements \cite{Smith2019}. These components are critical for replicating the subtle nuances of natural movements and expressions.
		\item Developments in miniature actuators and motors have allowed for more intricate and compact designs \cite{Brown2020}, enabling the creation of smaller, more detailed animatronic figures without compromising on performance.
	\end{itemize}
	\item \textbf{Sophisticated Control Systems}:
	\begin{itemize}
		\item The integration of advanced control systems, such as microcontrollers and digital signal processors, has greatly improved the coordination and synchronization of animatronic movements. These systems process input from sensors and execute complex motion sequences with high precision.
		\item Control algorithms utilizing artificial intelligence, including neural networks and reinforcement learning, have enhanced the adaptability and learning capabilities of animatronics \cite{mnih2015human, russell2016artificial}. These algorithms allow animatronic creatures to adjust their movements in real-time based on environmental feedback, leading to more dynamic and realistic interactions.
	\end{itemize}
	\item \textbf{Computational Design and Simulation Tools}
	\begin{itemize}
		\item The use of CAD (Computer-Aided Design) software has become indispensable in the design of animatronic figures. These tools allow for the creation of detailed 3D models and simulations, enabling designers to visualize and test mechanical systems before physical production \cite{zhang2012}.
		\item CAE (Computer-Aided Engineering) tools facilitate the analysis of mechanical stress, fluid dynamics, and thermal properties, ensuring that animatronic designs are robust and capable of withstanding the demands of filming. The adoption of ANCF for simulating flexible multibody dynamics has significantly improved the accuracy of these analyses, allowing for more detailed and realistic motion predictions.
	\end{itemize}
\end{itemize}

\subsection{Improving Animatronic Motion Dynamics}
The aforementioned advancements in mechatronics have led to significant improvements in the functionality and realism of animatronic creatures. The integration of flexible multibody dynamics, particularly through the use of ANCF, has facilitated smoother and more sophisticated movements by providing a more accurate representation of complex deformations and interactions. These technologies have enabled animatronics to perform lifelike actions that were previously unattainable \cite{Nada2023}. These technologies have facilitated smoother and more sophisticated movements through various means:

\begin{itemize}
	\item \textbf{Precision and Control}:
	\begin{itemize}
		\item High-precision sensors and actuators enable finer control over animatronic movements, allowing for smooth transitions and subtle gestures. This precision is essential for replicating the natural motions of living organisms, such as the fluid motion of a fish or the intricate facial expressions of a humanoid character.
		\item The implementation of ANCF in motion control systems has provided a more accurate and robust framework for managing the complex dynamics of flexible animatronic components \cite{El-Hussieny2023DeepCNN}.
	\end{itemize}
	\item \textbf{Real-Time Adaptability}:
	\begin{itemize}
		\item The integration of AI-driven control systems has endowed animatronics with the ability to adapt their movements in real-time. By processing sensory input and adjusting motion patterns on-the-fly, these systems can react dynamically to changes in the environment, resulting in more lifelike and believable performances \cite{russell2016artificial, El-Hussieny2023DeepCNN}.
		\item ANCF-based simulations allow for real-time adjustments to the movement of flexible components, enhancing the interactive capabilities of animatronic systems \cite{Bayoumy2014,nada2024development}.
	\end{itemize}
	\item \textbf{Complex Motion Sequences}:
	\begin{itemize}
		\item Advanced computational tools and control algorithms have enabled the programming of complex motion sequences that were previously unattainable \cite{Nada2024embedded}. Animatronics can now perform intricate actions, such as coordinated group behaviors or synchronized movements, which enhance the overall realism and immersion of cinematic scenes.
		\item The use of flexible multibody dynamics through ANCF has improved the ability to simulate and execute these complex sequences with high fidelity \cite{Bozorgmehri2019,Shabana15_1,Zhang2018}.
	\end{itemize}
	\item \textbf{Improved Durability and Flexibility}:
	\begin{itemize}
		\item The use of advanced materials has increased the durability and flexibility of animatronic components, allowing them to withstand rigorous use without compromising on performance. Flexible skin materials and robust joint designs contribute to more fluid and natural movements, further enhancing the realism of animatronic creatures \cite{Abdelsalam2021,Erukainure2022,Abdelsalam2023}.
		\item ANCF within the Multibody system dynamics has played a crucial role in modeling the interactions and deformations of these rigid and flexible components, ensuring that they perform reliably under various conditions \cite{shabana2023state,Nada2024}.
	\end{itemize}
\end{itemize}

\section{Case Studies of Mechatronics in Film-making}
\begin{figure}[tb]
	\begin{center}
		\begin{minipage}[tb]{0.75\textwidth}
			\includegraphics[width=0.90\textwidth]
			{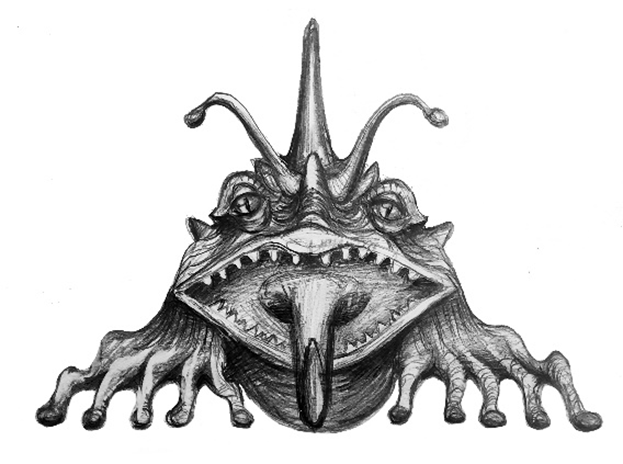}
			\caption{A Concept Sketch Showing the Front View of the Worrt Character}
			\label{Fig: Worrt1}
		\end{minipage}
	\end{center}
\end{figure}
Following the completion of the research study on the advancements in mechatronics technology and their noteworthy influence on the techniques employed in the implementation and construction of creatures and characters in movies. In this section, the Worrt character is introduced as a case study to explore the application of animatronics and CGI (Computer-Generated Imagery) in film-making. Worrt character provides a sufficient and in-depth examination of the techniques involved.

Worrt is a large, voracious amphibian known for its insatiable appetite, consuming anything that crosses its path. One of Worrt's defining physical features is its long, elastic tongue. This appendage is highly flexible, capable of moving in all directions, making it particularly effective in capturing prey. The tongue's mobility and reach are central to Worrt's feeding habits and are essential considerations in its design and animation.

The primary distinction between the shark, as seen in movies like "Deep Blue Sea," and the Worrt character lies in their design and the methods required to bring them to life on screen. The shark, a familiar and realistic creature, was primarily created using a blend of advanced animatronics and CGI to ensure it appeared lifelike and menacing in its aquatic environment. In contrast, Worrt, being a fictional and imaginative amphibian, presents unique challenges that necessitate a combination of digital modeling and possibly more intricate animatronics due to its fantastical design elements.

The advancements in mechatronics outlined in section (5) provide significant opportunities for modeling, actuating, and imagining the Worrt character in filmmaking. Here's some suggestion of each item can be applied as described in the following discussion.

\subsection*{Continuum modeling}
Worrt's design, with its highly flexible tongue and limbs, can be accurately modeled using the ANCF approach, which captures both the overall motion and the intricate local deformations of the body. This is particularly important given the Worrt's amphibian nature, which requires the simulation of soft, elastic movements alongside more rigid structural motions. ANCF provides a framework where the mass matrix remains constant, even in the presence of large deformations. This is crucial for achieving stable and efficient simulations, especially in dynamic scenes where the character interacts with its environment in real-time. For Worrt, whose movements might involve rapid and substantial changes in body posture, maintaining a constant mass matrix helps in accurately capturing the inertia effects without introducing unnecessary computational complexity. Moreover, the formulation of elastic forces within the ANCF allows for a more realistic representation of Worrt's material properties, whether those involve soft tissue or more rigid sections of its body. This is achieved through continuum mechanics approaches, which are integral to the ANCF method. By defining elastic forces in a manner that accounts for the Worrt's unique anatomy, ANCF can simulate both the gross and subtle deformations that would occur as the character moves or interacts with objects within its environment. Given these features, the ANCF approach presents a highly suitable method for both the CGI modeling and potential physical construction of Worrt using advanced mechatronics. For physical construction, ANCF's ability to accurately model large deformations and complex geometries can inform the design of animatronic components, ensuring they replicate the intended movements and flexibility of the CGI model.

\subsection*{Computational Design and Simulation Tools}
\begin{figure}[tb]
	\noindent 
	\begin{minipage}[tb]{0.33\textwidth}
		\begin{center}
			\includegraphics[width=0.92\textwidth]
			{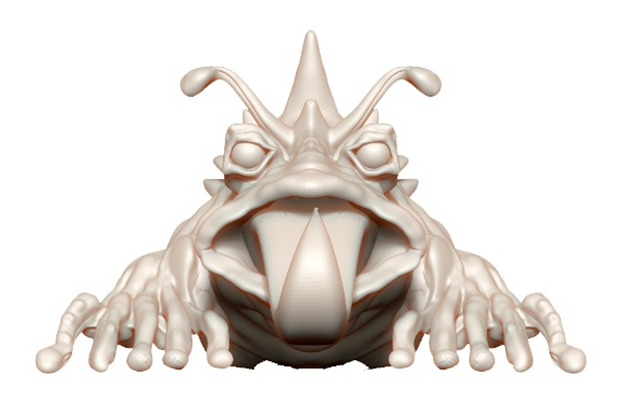}
\caption{Front-View of the Worrt}
\label{Fig: Worrt2}
		\end{center}
	\end{minipage}\noindent 
	\begin{minipage}[tb]{0.33\textwidth}
		\begin{center}
			\includegraphics[width=0.99\textwidth]
			{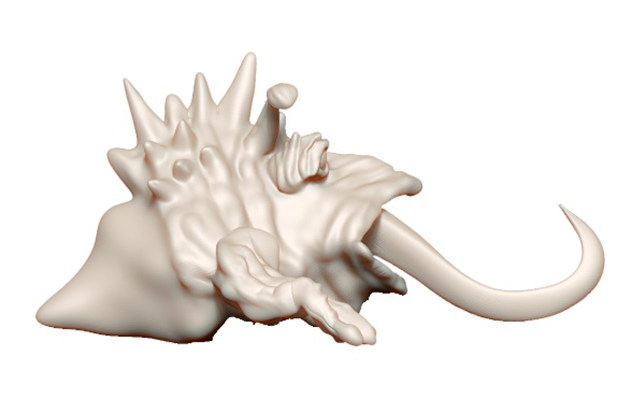}
\caption{Side-View of the Worrt}
\label{Fig: Worrt3}
		\end{center}
	\end{minipage}
	\begin{minipage}[tb]{0.33\textwidth}
		\begin{center}
			\includegraphics[width=0.99\textwidth]
			{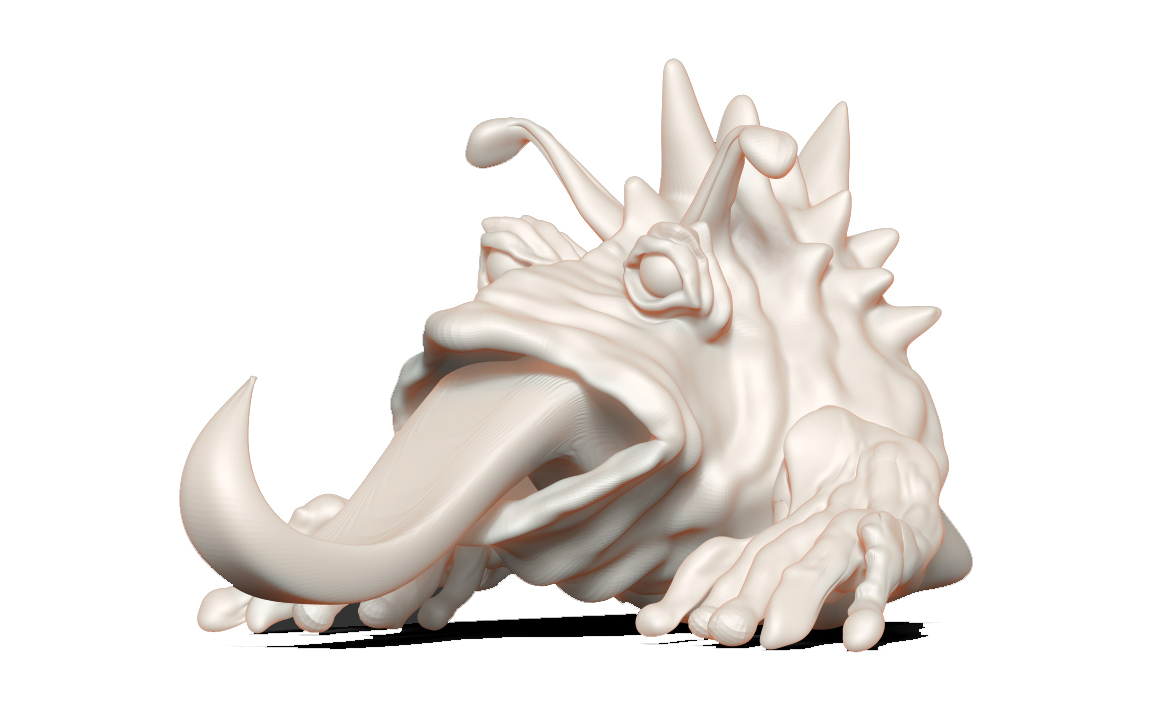}
\caption{Perspective-View of the Worrt}
\label{Fig: Worrt4}
		\end{center}
	\end{minipage}\newline
\end{figure}

Creating a CGI model of Worrt involves several important steps. First, the character design begins with 2D sketches that offer different perspectives top, front, side, and back views that capturing Worrt’s key features and serving as the blueprint for the digital model, see Fig. (\ref{Fig: Worrt1}). These illustrations laid the groundwork for further development by defining Worrt's form, anatomical details, and potential range of movements, see Figs.(\ref{Fig: Worrt2}-\ref{Fig: Worrt4}). Second, ZBrush software is used to translate these 2D sketches into a highly detailed 3D model. Known for its powerful sculpting capabilities, ZBrush allows artists to create a realistic and dynamic representation of Worrt, a step that is crucial for refining the character’s appearance and ensuring it can be animated effectively. Lastly, the completed 3D model undergoes rigging, where a digital skeleton is added to facilitate movement. This rigged model can then be animated to perform various actions, such as the characteristic movement of Worrt's elastic tongue, bringing the character to life on screen.

\subsection*{Advanced Materials and Manufacturing Techniques}
First, the development of new flexible polymers and lightweight composites enhances the durability and flexibility of the Worrt character’s animatronic components. This allows for more lifelike skin textures and fluid movements, which are critical for accurately depicting Worrt’s unique physical features, such as its elastic tongue. Additionally, innovations in 3D printing enable rapid prototyping, allowing filmmakers to quickly iterate and refine the design of the Worrt character, ensuring that it performs optimally in various scenes.

\subsection*{Enhanced Electronics and Actuation Systems}
Second, the integration of modern animatronics with advanced electronics, including high-precision sensors and actuators, allows the Worrt character to achieve more accurate and responsive movements. These components are essential for replicating the subtle nuances of Worrt’s natural behaviors, such as the quick flicks of its tongue or its subtle body shifts. Moreover, developments in miniature actuators and motors enable the creation of a compact and intricate design for the Worrt character without sacrificing performance, making it easier to incorporate Worrt into various scenes seamlessly.

\subsection*{Sophisticated Control Systems}
Lastly, the use of advanced control systems, including microcontrollers and AI-driven algorithms, greatly improves the coordination and synchronization of Worrt’s movements. For example, AI algorithms can allow Worrt to adjust its movements in real-time based on environmental feedback, leading to more dynamic and lifelike interactions. This is particularly useful for scenes where Worrt interacts with live-action elements or other CGI characters. Moreover, the adoption of Absolute Nodal Coordinate Formulation (ANCF) in motion control systems offers a more accurate framework for managing the complex dynamics of Worrt’s flexible components, ensuring that its movements are smooth and realistic.

\begin{figure}[!p!t]
	\centering
			\includegraphics[width=0.7\textwidth]
			{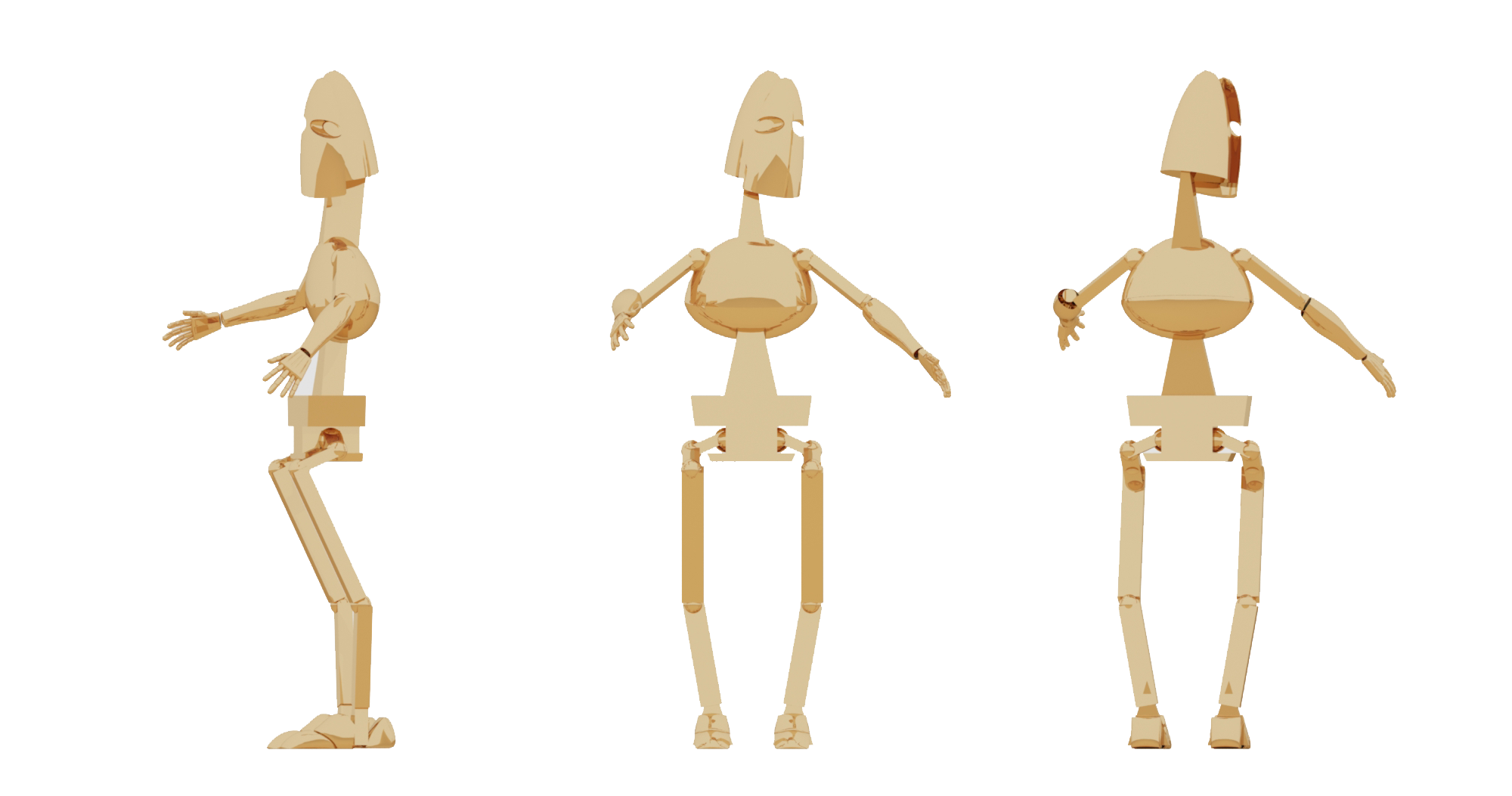}
			\caption{Humanoid animatronic character design featuring segmented joints. The figure illustrates potential locations for actuation (in the arms, legs, and torso) using servo motors or linear actuators, and joints where sensors such as encoders could provide feedback for precise motion control. This character could be realized using mechatronic systems integrating advanced control algorithms for lifelike movements.}
			\label{Fig:humaniod}
\end{figure}
Another case study of is a simple creature in Fig. \ref{Fig:humaniod} representing a humanoid robot with segmented joints, which could be implemented in animatronics for movies. This animatronic setup could deliver realistic, repeatable performances of this character, blending mechanical design, electronics, and advanced control systems for dynamic performance in films. Realizing this character with mechatronic systems involves the integration of several components:

\subsection*{Mechanical Structure and Actuation:}
The segmented body and limbs of this character suggest the use of servo motors or linear actuators at each joint, providing precise control of movements such as bending and rotating. The arms, legs, and torso can be driven by DC or stepper motors depending on the required precision and load. Gears and linkages can be employed to achieve natural and smooth movement transitions.

\subsection*{Sensors:}
Position sensors such as encoders or potentiometers can be embedded in the joints to measure angles and provide feedback for control. For more fluid, naturalistic movement, IMUs (Inertial Measurement Units) can monitor the character's posture in real-time and help maintain balance. Pressure or tactile sensors could be used in the hands and feet to simulate gripping or walking actions.

\subsection*{Control Systems:}
A PID controller would provide smooth motion for basic tasks. However, more complex movements like walking or gesturing may require Model Predictive Control (MPC) or advanced kinematic algorithms to handle multiple degrees of freedom simultaneously. An onboard microcontroller (such as an Arduino or Raspberry Pi) could handle real-time control, interfacing with all actuators and sensors.

\subsection*{Power Supply:}
- Given the potentially complex and continuous movements, a battery-powered or external power source with careful management of energy distribution would be critical to ensure operational efficiency.

\subsection*{Motion Capture and Programming:}
To create lifelike movements, motion capture technology can be used to record human actions that are then translated into commands for the character’s joints. Pre-programmed sequences can also allow for complex animated routines, especially for facial expressions if the design incorporates finer details like eyes, mouth, or subtle head movements.

\section{The Future of Creature Animation}
The future of creature animation is being reshaped by advancements in mechatronics technologies, particularly in Hollywood and the broader entertainment industry. Mechatronics, which integrates mechanical engineering with electronics and computer control, is revolutionizing how film-makers create and bring creatures to life on screen.

1) Realism and Interaction:
Mechatronics allows for much more realistic creature movement and interaction with the environment and human actors. This realism provides immersion and puts audiences more in emotional connection with characters.

2) Enhanced Articulation :
Mechatronics Search and Development enhances conventional methods of animatronics as a means of achieving complicated and fluent motions. It enunciates body language, facial expression, and even the minutest features like moving scales or fur in a very realistic way.

3)Integration with CGI and Other Technologies :
 The future likely involves deeper integration of mechatronics with other technologies like CGI (Computer-Generated Imagery) and AI (Artificial Intelligence) to further enhance realism and interactivity.

	\section{Conclusion}
The latest advancements in mechatronic technologies have significantly enhanced the capabilities of animatronic creatures in film-making. By integrating advanced materials, electronics, control systems, and computational tools, modern animatronics can achieve smoother, more sophisticated, and lifelike movements. These technological innovations continue to push the boundaries of what is possible in animatronic design, leading to increasingly immersive and believable cinematic experiences. Future research and development in this field will undoubtedly yield even more impressive advancements, further revolutionizing the role of animatronics in the film industry.

In the filmmaking industry, the Absolute Nodal Coordinate Formulation (ANCF) is a powerful tool for creating realistic animations, particularly in special effects and character animation. ANCF excels at simulating the complex dynamics and deformations of flexible structures such as ropes, cables, hair, and clothing. By enabling detailed modeling of large deformations and dynamic interactions, ANCF allows animators to produce lifelike movements that respond naturally to characters' actions and environmental forces. This results in enhanced realism, reduces the need for manual adjustments, and offers versatility in animating a wide range of flexible objects, making ANCF an invaluable asset in the creation of high-quality cinematic experiences.

the ANCF approach offers a robust framework for animating and modeling Worrt by accurately capturing large deformations, ensuring computational efficiency, and representing complex material behaviors. This would be particularly useful not only in CGI animation but also in the design of animatronic systems that might be used for live-action or interactive elements within a production.

The advancements listed in the paper not only allow for the creation of a highly detailed and realistic CGI model of Worrt but also open up possibilities for constructing a physical animatronic version that can perform in a variety of filmmaking scenarios. The combination of these cutting-edge technologies ensures that the Worrt character can be brought to life with a high degree of realism and believability, enhancing the overall cinematic experience.

	\section*{Conflict of interest}
	The authors declare that they have no conflict of interest.
	
	\bibliographystyle{model1-num-names}
	\bibliography{ANIMATRONICS_2024_04}
	
\end{document}